# Hours-of-service-aware siting of charging and battery-swapping stations for long-haul electric trucks under adoption uncertainty

Elnaz Irannezhad*[a], Jia Guo[a]

[a]Research Centre for Integrated Transport Innovations (rCITI), School of Civil and Environmental Engineering, University of New South Wales, Sydney, New South Wales, Australia, 2052

* Corresponding author's email: e.irannezhad@unsw.edu.au

## Abstract

Planning en-route charging and battery-swapping infrastructure for long-haul battery-electric trucks (BETs) requires models that reflect how trucks actually operate. This paper develops a mixed-integer programming framework that jointly sites charging or swapping stations and schedules each truck's charging, swapping and mandatory driver rest, so that charging time overlaps with regulated rest instead of being added to it. Energy use is derived segment by segment from road terrain with a tractive-force model, and the truck battery is modelled as a set of independently swappable packs. Staged investment under uncertain BET adoption is formulated as a multistage stochastic program with Markovian demand and solved by stochastic dual dynamic integer programming (SDDiP) with Lagrangian cuts; we show that the Lagrangian multipliers can be bounded by each station's annualised cost without weakening the cuts. Applied to twelve freight corridors on Australia's East Coast, the algorithm jointly optimises charging and battery swapping events as well as mandatory break events. A ±30% change in adoption alters the final charge-only network by only −12% to +13% of stations, with over 95% of stations built by the second stage. Hedging against uncertainty mainly changes which sites are chosen (71% overlap with a deterministic rolling-horizon model), not when they are built.

***Keywords:*** Battery-electric trucks; Charging infrastructure; Hours of service; Multistage stochastic programming; Stochastic dual dynamic integer programming; Driver rest regulations

# 1 Introduction

Battery electric trucks (BETs) are widely regarded as a key pathway to decarbonising road freight, and attention is shifting from urban and regional operations to long-haul corridors (Balke et al., 2024). Their operational feasibility depends not only on battery capacity but also on how charging events are represented, how limited station capacity is allocated and how charging interacts with vehicle schedules. Unlike diesel refuelling, battery replenishment may require substantial dwell time and is affected by the nonlinear charging profile, available infrastructure and uncertain demand. Infrastructure planning must therefore connect spatial deployment decisions with operational constraints, including delivery schedules and mandatory driver rest periods.

Return-to-base operations, in which trucks complete short routes and recharge at a depot, are a plausible early deployment pathway. Long-haul operations require a different planning problem because en-route charging or battery swapping must be coordinated with route geometry, energy use, station capacity and driver regulations. A useful planning model should therefore identify infrastructure that is feasible for representative freight movements while making its assumptions about demand, technology and regulation explicit.

Distance-based station placement alone does not guarantee an economically or operationally feasible network. Existing heavy-vehicle rest areas, passenger electric vehicle (EV) charging sites and diesel refuelling locations may offer different combinations of land, access, amenities and electrical infrastructure. Co-locating replenishment facilities with rest areas may reduce the incremental delay associated with charging when the charging event can overlap with a required break. This is particularly advantageous, as it aligns charging dwell time with regulated driver rest periods, thereby minimising productivity impacts. Where feasible, expansion of existing EV charging infrastructure may also reduce marginal upgrade costs, though dedicated high-capacity connections or on-site energy management solutions for heavy vehicles may still be required. Finally, existing diesel refuelling sites along major freight corridors present strong candidates for early-stage retrofitting due to their established heavy-vehicle access, strategic locations, and supporting amenities; however, the significantly higher power requirements of heavy-duty charging necessitate careful assessment of electrical capacity, civil works, and site layout constraints. Importantly, infrastructure planning should adopt a network-level optimisation framework accounting for truck charging demands and prioritising strategically located stations at freight corridor intersections to maximise utilisation and reduce redundant capital expenditure (CAPEX) investments. Capacity and station siting design must be informed by freight volume projections and peak demand modelling to ensure operational reliability while maintaining cost efficiency.

This paper develops a staged planning framework that combines uncertain BET demand, candidate-site feasibility, station capacity and route-level charging or swapping schedules. The framework considers demand stages associated with increasing BET penetration and identifies station locations and capacity requirements as demand evolves. It also evaluates whether regulated rest periods can absorb some replenishment time. BET charging demand on each corridor is derived from observed traffic volumes at several adoption levels, and its growth is represented by a Markov chain over low, mid and high adoption branches. These stochastic demands feed a cost–delay optimisation model that determines which candidate sites to open at each adoption stage, how many chargers or swap batteries each site requires, and how each truck combines charging or swapping with the mandatory short breaks and long rests of its driver. The result is a staged deployment plan whose site choices hedge against the range of adoption paths.

The literature on BET charging infrastructure heavily focuses on mathematical optimisation methods to address facility location or vehicle routing problems. Whitehead et al. (2021) argue that current optimisation models of this nature function well for strategic planning. However, they have difficulty in representing the dynamic complexity of actual operations. The early BET studies have simplified the technical and operational constraints. For example, numerous optimisation models adopt the charging time and battery state-of-charge (SoC) as a linear relationship. The lithium-ion battery is charged following a non-linear profile, and the speed of charging is reduced when the battery is almost full, which is the tapering phase. When the non-linear behaviour of the battery is ignored, it is possible that "optimistic" operational schedules will appear that cannot be realised in practice (Montoya et al., 2017).

Furthermore, early studies often relied on other simplified assumptions, such as the requirement to always fully recharge the battery. Although overnight depot charging remains the most common strategy for fleet operations to avoid downtime, opportunity charging through battery swapping or fast charging can extend the daily operating range. Furthermore, the separation of charger deployment from vehicle scheduling usually results in the inefficient use of capital (Bertucci et al., 2024). It becomes particularly important for heavy vehicle operations, because heavy vehicles must take mandatory rests and follow the fatigue management schemes set by heavy vehicle regulators. The mandatory truck driver rest duration can be utilised for charging or battery swapping purposes.

Three questions motivate this study. First, how much charging or swapping infrastructure do long-haul freight corridors require when energy use is derived from terrain, charging follows a non-linear curve, and charging or swapping is scheduled jointly with mandatory driver rest? Second, how sensitive is that requirement to the balance between capital cost and driver delay, and does the sensitivity differ by corridor? Third, how much does planning under adoption uncertainty, rather than for a single forecast, change which stations are built and when?

To answer these questions, this study makes five contributions. First, unlike a flat, distance-proportional approximation of energy use, the proposed model derives energy consumption directly from the road corridor's vertical alignment. The road centreline for each route is discretised into 1-kilometre segments, each of which is assigned a terrain-derived grade. A physics-based tractive-force model is then applied to each segment to determine whether the truck draws energy from the battery (level or uphill travel) or returns energy to the battery through regenerative braking (sufficiently steep descents), and a constant auxiliary electrical load is added on top of traction demand. Segment-level energy values are summed to obtain cumulative energy consumption between the origin and destination of each route, and between consecutive candidate charging/swapping facilities along the route.

Second, the model jointly sites stations and schedules charging or swapping together with mandatory driver rest. Charging and rest are allowed to overlap, so charging adds to trip time only when it outlasts the rest the driver must take anyway. Earlier studies schedule charging and rest along a given route with given stations (Bai et al., 2023; Wan et al., 2024; Zähringer et al., 2024; Kahlert et al., 2026); here station locations are decision variables. We implement a stylised two-tier rule (a 10-minute short break after at most 2 hours of continuous driving and a 30-minute long rest after at most 5 hours of cumulative driving) that requires more frequent rest than the minimum under Australia's Standard Hours (Section 4). Other two-tier regimes can be represented by only changing four parameters.

Third, this study considers multi-pack batteries on trucks instead of one battery. This introduces a more realistic assumption where each battery is independently swappable. Most existing studies target passenger cars and last-mile delivery vans, which carry a single battery pack, so the battery has not been modelled as an indexed set of packs. Modular heavy-truck swapping systems already allow one to three standardised packs per truck (Barnard, 2026). In this study, we assume each truck is equipped with a fixed number of battery modules. In a modular battery-swapping system, the number of packs exchanged during a single swapping event may vary, such that either a single pack or multiple packs can be replaced. By contrast, for charging, the onboard charging system distributes power across the individual battery modules, enabling simultaneous charging of all batteries.

Fourth, staged investment under uncertain adoption is formulated as a multistage stochastic program with Markovian demand and solved by stochastic dual dynamic integer programming (SDDiP), building on Zou et al. (2019). At each adoption stage, the policy decides which stations to open, given the stations already built and the adoption level observed so far; an opened station is never closed. Because station costs are discounted, a station that is not yet needed is cheaper to build later, and the stochastic policy weighs this against the risk of having to add stations if adoption turns out high. Methodologically, multistage stochastic integer programs of the kind formulated here are increasingly solved with SDDiP (Zou et al., 2019), which has been applied to power-system capacity expansion (Lara et al., 2020). Stagewise-dependent uncertainty is commonly handled by modelling the random process as a Markov chain and keeping a separate cost-to-go approximation for each Markov state (Dowson and Kapelevich, 2021; Löhndorf and Shapiro, 2019). We adopt this approach for BET adoption. We also show that the Lagrangian multipliers used to generate SDDiP cuts can be bounded by each station's annualised cost without weakening the cuts (Proposition 2), and that the cuts remain valid when dual iterations or MILP solves are stopped early (Proposition 1). Fifth, a route's peak-hour flow can be split into several parallel sub-flows, each free to use different stations, so that peak demand is served jointly by multiple capacity-constrained stations rather than by a single facility.

Section 2 provides a review of the literature and research gap. Section 3 introduces the methods and model formulation. Section 4 presents the case study, and Section 5 presents the results. Section 6 discusses the results, and Section 7 concludes the study and outlines study limitations and future research directions.

# 2 Literature Review

This review is organised around four linked modelling layers: evidence on BET deployment, energy-consumption estimation, charging behaviour and scheduling, and infrastructure placement under capacity and uncertainty. We identify which assumptions are inherited when operational models are used to inform strategic station investment. The review concludes by specifying the modelling gaps addressed in this paper and the aspects that remain outside its scope.

## 2.1 Operational Feasibility and Deployment Evidence

Field evidence indicates that BET performance depends strongly on duty cycle, battery capacity, ambient conditions and the maturity of the supporting charging network. Hovi et al. (2019) for example, reported positive user experiences among seven Norwegian pilot operators, but also identified range limitations under high loads and cold conditions, which have led companies to re-route operations and be cautious about vehicle scheduling during peak hours. These findings support modelling BET operations as context-dependent rather than assuming that one vehicle and infrastructure configuration is suitable for every freight task.

Economic studies reach a similar conclusion. Vijayagopal and Rousseau (2021) showed that the relative competitiveness of BETs depends on vehicle class, battery cost, electricity price and annual utilisation. In particular, a large battery can be economically inefficient when the duty cycle does not use its range. These results motivate explicit battery and duty-cycle assumptions in infrastructure studies, although the present paper does not perform a full total-cost-of-ownership comparison.

At the network level, Liimatainen et al. (2019) found that electrification potential varies by commodity, vehicle class and charging strategy. Their results demonstrate that the infrastructure network can be a binding condition for long-haul electrification. They suggested that, at current battery technology, BETs can cover around 52% of road transport tasks in Switzerland and 50% in Finland; with higher-capacity batteries and fast-charging infrastructure, coverage rises to 93% and 81% respectively. This gap between current and achievable coverage underscores the importance of the charging network itself, rather than the vehicle alone, as the binding constraint on electrification. The same study identifies medium-duty trucks (under 26 tonnes) and some heavy-duty rigid trucks (26-40 tonnes) as having the strongest commercial case for early electrification, informing the vehicle classes represented in this study's model.

These studies motivate two modelling requirements: energy and range estimates should reflect route conditions, and infrastructure should be evaluated against the operating patterns it is intended to serve. The following subsections review how these requirements have been addressed and where simplifications remain.

## 2.2 Energy Consumption Modelling

Energy-consumption estimates affect feasible route spacing, charging requirements and the reliability of the resulting schedule. The literature generally uses either physics-based longitudinal models or data-driven models. The former are transparent and can be applied where telematics data are limited, whereas the latter can capture empirical interactions if sufficiently detailed training data are available.

Physics-based models use vehicle-dynamics relationships to estimate the power required to overcome aerodynamic drag, rolling resistance, gravitational effects and inertia. Goeke and Schneider (2015) incorporated such relationships into an electric-vehicle routing problem, including payload effects. Tong et al. (2021) likewise showed that corridor-level energy use can vary materially with road grade and speed. These studies support using route-specific energy estimates when station spacing is a central decision.

Physics-based models are explainable but require detailed parameters, such as drag coefficients and rolling resistance, that are not always available or accurate under real-world conditions. Data-driven approaches tackle this issue by regressing the observed energy consumption over variables such as speed, road grade and vehicle weight, using real-world empirical data. As the availability of BET telematics data increases, researchers move beyond classical statistical models and apply machine learning methods to unravel the nonlinear relationships between parameters (e.g. see, (Gonzalez et al., 2023)). Such models may be valuable for calibration, but detailed second-by-second BET data remain difficult to obtain, especially for emerging long-haul fleets.

## 2.3 Charging Behaviour and Non-linear State-of-Charge Dynamics

Once energy use is estimated, charging duration must be represented consistently with battery behaviour. A significant advance in this literature is the treatment of the relationship between charging time and battery SoC as non-linear rather than linear. Montoya et al. (2017) showed that nonlinear charging functions can materially affect electric-vehicle routing decisions. A constant-current/constant-voltage (CC-CV) profile generally produces a tapering rate at high SoC, so a linear charging-time approximation may underestimate dwell time. Piecewise-linear approximations provide a tractable way to embed this relationship in a mixed-integer model, provided that their breakpoints and approximation error are reported. Löbel et al. (2025) further discussed the limitations of coarse linear interpolation for nonlinear battery behaviour. Their findings highlight the need to document breakpoint selection and to check whether the approximation is conservative. Together, these two studies establish that charging-time realism, not just energy realism, is a prerequisite for credible station-sizing and scheduling models.

## 2.4 Charging Scheduling and Grid Interaction

With energy and charging time specified, a further literature stream considers operational scheduling and grid interaction. Bertucci et al. (2024) studied charging infrastructure placement jointly with the charging capacity in a logistics distribution network. Walz and Rudion (2024) demonstrated that charging policies affect aggregate load profiles on the electricity grid, while Balke et al. (2024) examined multi-site charging and synchronisation with rest periods, reducing peak-hour waiting time and spreading charging load. Liimatainen et al. (2019) simulated four charging-strategy scenarios, including centralised night charging and high-power on-road charging, and found that the share of routes coverable by BETs nearly doubles when a minimum charging time of 2 hours and charging power of at least 250 kW are available on-road, compared with relying on night charging alone, reinforcing the importance of high-power en-route infrastructure for range coverage. These studies suggest that vehicle schedules, charger capacity and grid impacts are interdependent. The present model captures the first two interactions but leaves electrical-network constraints outside the formulation and suggests it for future research direction.

Shared facilities also create congestion when truck arrivals overlap. Bai et al. (2023) modelled a coordinated charging mechanism and queue information sharing between charging stations and drivers. They modelled a scenario in which stations provide waiting-time estimates on a first-come, first-served basis and trucks dynamically re-optimise charging decisions en-route. Their results indicate that capacity planning based only on individual vehicle energy needs may be insufficient. In contrast, the present model represents peak concurrent demand through capacity constraints but does not model stochastic queues or first-come, first-served service explicitly.

## 2.5 Charging and Battery-Swapping Infrastructure Placement

The most closely related literature concerns the location and sizing of charging or battery-swapping infrastructure. These studies typically use facility-location, vehicle-routing or network-design formulations to balance coverage, service quality and investment cost. Their assumptions differ in whether station capacity, route-level scheduling, grid constraints and uncertainty are represented.

Whitehead et al. (2021) used a facility-location approach to examine public charging requirements for short-haul electric trucks in Queensland. Their results illustrate how public charging can complement depot charging, while also highlighting the limitations of coverage models that omit station capacity. Speth et al. (2025) incorporated capacity constraints informed by queueing considerations, showing why station layouts that ignore service capacity may be operationally infeasible. Hurtado-Beltran et al. (2021) used GIS-based network analysis to assess electrification potential across the U.S. highway system, and proposed that existing truck-stop infrastructure can provide a useful basis for charging deployment, although electrical and site constraints remain important. Deng et al. (2024) incorporated environmental objectives into station planning and demonstrated that a small number of well-located sites can cover a large share of freight flow in a particular study area.

Other studies consider battery-swapping systems and the management of batteries at swap stations. Outside the truck context, several studies address battery-delivery and battery-swapping station siting for passenger electric vehicles, without a truck or charging-network focus (e.g., (Liu et al., 2019; Yang and Sun, 2015; Yang et al., 2024)). Within trucking, Wang et al. (2024) jointly modelled station design and the charging of depleted batteries, while Ye (2024) considered charging infrastructure and grid supply for heavy-duty drayage.

A smaller group of studies incorporates driver regulations or dynamic charging into operational scheduling. Zähringer et al. (2024), Wan et al. (2024), and Kahlert et al. (2026) represent aspects of charging and rest scheduling, but treat station locations as given or simplify the regulatory process. Wang et al. (2025) compare swapping and fast charging in a strategic framework, but do not use the same route-specific station-siting formulation.

In operations research, en-route station siting is usually framed as a flow-refuelling location problem. An origin–destination flow is served only if the stations along its path allow the trip to be completed within the vehicle range (Capar et al., 2013; Kuby and Lim, 2005). Later work added Benders Decomposition algorithm for plug-in hybrid vehicles (Arslan and Karaşan, 2016), route flexibility under full coverage (Kınay et al., 2021), multi-period deployment (Chung and Kwon, 2015), multi-period capacitated planning with demand dynamics (Zhang et al., 2017) and non-linear charging times with heterogeneous chargers (Park and Lee, 2024). For battery swapping, Mak et al. (2013) located swapping stations for passenger EVs under uncertain demand using robust optimisation. These models treat a trip as a sequence of range-feasible legs; none of them schedules the driver's mandatory rest, which is a defining operational constraint of long-haul trucking.

Truck driver scheduling under working-hour rules is itself a well-studied problem (Goel, 2009). Goel (2012b) formulated the Australian truck driver scheduling problem, whose sliding-window rules limit work to at most 12 hours and require at least 7 hours of rest in any 24-hour period. Similarly, (Goel, 2010, 2012a; Goel, 2012c) formulated the truck driver schedules for the European Union, Canada and the US. These studies treat vehicle range as unlimited, whereas the present paper models the interaction between rest and energy replenishment.

Collectively, these studies motivate a model that links station location, capacity and regulated rest scheduling. The present study is the first to address everyone of these features jointly, so the contribution is stated as an incremental integration of established modelling elements.

## 2.6 Research Gaps and Study Contributions

The reviewed literature reveals a recurring separation between operational realism and strategic infrastructure planning. Energy models are not always connected to station siting; scheduling studies often treat station locations as fixed; and infrastructure studies frequently use static demand and simplified battery or rest representations. Studies that model terrain-based energy physics in detail (e.g. (Goeke and Schneider, 2015; Tong et al., 2021)) are not connected to siting or scheduling decisions; siting and scheduling studies that address driver rest regulations (e.g. (Kahlert et al., 2026; Wan et al., 2024; Zähringer et al., 2024)) either treat station locations as fixed or simplify the rest rules; and the placement and swapping studies reviewed treat the truck battery as a single pack and the infrastructure network as a static, single-stage design problem.

This paper combines several of these elements in a single staged planning framework. Table 1 summarises the gaps, their planning relevance and the corresponding modelling choices. Taken together, these five contributions integrate the energy, charging-behaviour, scheduling and network-placement layers reviewed above into a single staged planning framework, closing the gap between the operational realism established by the energy-consumption, charging-behaviour and scheduling literature and the strategic, investment-timing decisions addressed by the facility-location and battery-swapping literature.

*Table 1: Research gaps identified in the literature, their importance, and this study's contributions.*

| Research gap | Why it matters | Contribution of this study |
|---|---|---|
| Energy consumption in siting/scheduling models is typically considered a flat, distance-proportional rate rather than derived from actual road grade and vehicle dynamics (e.g. (Kahlert et al., 2026; Tong et al., 2021)). | Specific energy consumption varies by nearly a factor of two with road grade ((Tong et al., 2021)), so flat-rate assumptions cause errors in station siting, battery sizing and charging-time estimates. | Discretises each route into fixed-length segments with terrain-derived grade and applies a physics-based tractive-force model per segment, including regenerative braking on descents, to obtain cumulative, corridor-specific energy profiles between stations. |
| To our knowledge, no siting model integrates mandatory driver HoS rest regulations into the joint charging-and-siting decision; related studies either schedule charging/rest on a fixed route without siting stations (e.g. (Wan et al., 2024; Zähringer et al., 2024)), or simplify the rest rules while treating stations as given (e.g. (Kahlert et al., 2026)). | Ignoring the interaction between rest breaks and charging produces networks that look efficient on paper but generate unnecessary trip delay in practice, since rest dwell-time is not used to offset charging time. | Integrates mandatory rest regulations directly into the charging/swapping and station-siting decision, aligning rest breaks with station placement to jointly minimise total trip delay. |
| Existing formulations treat the truck battery as a single pack, an assumption inherited from passenger-EV and last-mile-van routing research (e.g. (Wang et al., 2024; Wang et al., 2025)); they also ignore the possibility of partial opportunity charging. | Heavy-duty trucks can carry multiple battery packs; a single-pack assumption rules out partial charging/swapping strategies and understates the operational flexibility available to operators. | Models multi-pack batteries as an independently swappable set, enabling realistic partial charging/swapping decisions. |
| Placement studies are largely static, single-stage designs sized against one | Building full capacity upfront risks stranded assets if demand | Applies a multi-stage SDDiP approach (building on (Zou et al., |
| demand forecast rather than a sequential investment problem under uncertain demand growth (e.g. (Deng et al., 2024; Speth et al., 2025; Whitehead et al., 2021; Ye, 2024)). | grows slower than expected, while under-building risks costly retrofits; static models price neither risk. | 2019)) to time station investment under uncertain demand growth, separating stations worth building now from those to defer, and discounting CAPEX to present value. |
| Siting/capacity models typically assign each route's peak flow to a single station rather than allowing several capacity-constrained stations to jointly serve shared peak-hour demand (e.g. Whitehead et al., 2021). | Single-station assignment understates congestion risk and misprices capacity investment when several stations could share peak load along a corridor. | Allows peak-hour flow on a route to be served by multiple, capacity-contended stations simultaneously, within each stage of the SDDiP framework. |

# 3 Models

This section presents the assumptions and mathematical formulation. Given candidate locations along each route, the proposed model determines the optimal siting of charging stations and battery-swapping stations by minimising the sum of (i) the total annualised fixed cost of station establishment and (ii) the monetised cost of driver delay, both in AUD per year, subject to HoS regulations and station capacity constraints. The stochastic charging demand component is then formulated as an SDDiP framework to identify the staged station sitings and required battery inventories and/or chargers at each station.

The principal decision variables are grouped as follows:

- Station deployment: (i) the candidate locations at which charging or swapping facilities are opened, (ii) the number of charging units installed at each deployed charging facility; (iii) the number of fully charged replacement battery modules at deployed battery swapping station; (iv) the truck sub-flow and facility visit along a route; and (v) if charging is selected, the charging duration and departure SoC of the battery module.
- Drivers' rest schedule: The driver rest decision at each facility visit: no break, a short break or a long rest.

Constraints include:

- Energy feasibility: the battery state must support travel between the origin, consecutive facilities and the destination.
- Facility capacity constraints:
    - Charging capacity: the number of trucks represented by the concurrent peak flow cannot exceed the available charging capacity.
    - Swapping capacity: the number of battery modules exchanged concurrently cannot exceed the available replacement battery inventory.
    - Demand growth: as the assumed BET share increases, the required service capacity may increase. Station deployment is irreversible in the current stochastic formulation, whereas charger counts and battery inventories are re-optimised at each stage unless additional monotonicity constraints are imposed.
- Drivers Hours of Service constraints:
    - A driver must take the assumed short break after no more than the specified maximum period of continuous driving.
    - A driver must take the assumed long rest after no more than the specified maximum cumulative driving period. The exact regulatory interpretation is documented in Section 4 and should be verified against the applicable jurisdiction.

The objective function is defined as follows:

- The first objective minimises the total annualised station establishment cost, where the upfront establishment cost is converted to an equivalent annual cost over the planning horizon using the capital recovery factor.

- The second objective minimises the total delay cost incurred by trucks, calculated as the product of the delay experienced by each truck and the corresponding driver value of time. The delay at each station is determined by the time required for battery charging or swapping and, where applicable, mandatory rest and break requirements under HoS regulations. Because mandatory rest is required whether or not the truck charges, this term includes dwell time that a diesel truck would also incur; what the model can reduce is the dwell time in excess of the mandatory rest.
- Both objectives are expressed in AUD per year (delay through the driver value of time), so they are combined by simple addition and the value of time acts as the weight on delay.

## 3.1 Assumptions

This study assumes that batteries charged en-route are not required to reach 100% state of charge (SoC), whereas batteries obtained through swapping are assumed to be fully charged (SoC = 100%). A minimum SoC of 10% is enforced throughout to mitigate range anxiety. Charging battery packs takes substantially longer than swapping them (the charging parameters used in the case study are given in Section 4); a swap can be completed in about 10 minutes, as demonstrated in the eHaul project (Technische Universität Berlin, 2026). Depleted batteries removed from a truck during a swap are subsequently recharged to 100% SoC at the station, rendering them available for future use by other trucks.

Assumptions are outlined in two categories:

- Data assumptions:
  - For each origin-destination pair, the input demand is represented by a peak-hour truck flow for the relevant planning stage. The model treats this aggregate flow as divisible across parallel sub-flows; it does not represent individual truck arrivals.
  - Travel speed is fixed by segment and travel time between consecutive facilities is deterministic. Congestion, weather-related speed variation and incidents are not modelled.
  - Energy consumption is calculated from tractive force as a function of grade, speed, distance, vehicle mass and the parameters in Table 3. All trucks are assigned the same values for these parameters within an experiment.
  - Each origin-destination pair follows a predetermined route and direction. Rerouting, detours to facilities and route choice are outside the present model.
- Operational assumptions:
  - When charging is selected at a visit, the battery modules on the representative truck are charged in parallel according to the adopted charging-curve approximation.
  - When swapping is selected, one or more battery modules may be exchanged in the same event. The numerical experiments assume a fixed ten-minute event regardless of the number of modules exchanged; this simplifying assumption requires validation for a specific swapping technology.
  - Charging duration is represented by a convex piecewise-linear approximation of the adopted CC-CV charge curve, rather than by a constant charging rate.
  - During travel, all battery modules discharge in parallel and are subject to the same segment-level energy requirement.
  - Partial charging is permitted. Arrival SoC is constrained by the minimum reserve, while a swapped module is assigned the assumed full SoC. The phrase “minimum SoC reserve” is used below in place of “charging anxiety.”
  - All battery modules are assumed to leave the origin at the specified initial SoC, which is set to 100% in the experiments.

## 3.2 Mathematical Formulation

This section presents mathematical formulation. Table 2 presents the notations used in this study.

*Table 2: Model notations.*

| | |
|---|---|
| Sets and indices: | |
| $R, r$ | Set and index of routes (origin-destination pairs) |
| $I, i$ | Set and index of candidate charging station locations |
| $I_r$ | Sequence of potential stations along route $r$, from the origin to destination |
| $o_r, d_r$ | Origin and destination of route $r$ |
| $B$ | Set of batteries on a truck |
| $\Gamma_r$ | Set of parallel sub-flows of route $r$ |
| $K$ | Number of breakpoints used to discretise the charging curve |
| $T, t$ | Set and index of planning stages |
| Demand parameters: | |
| $f_r$ | Total BET flow per day on route $r$ |
| $f_r^{peak}$ | Concurrent peak-hour BET flow on route $r$, as opposed to $f_r$'s daily total |
| $d_{r,i,i+1}$ | Distance between consecutive locations $i$ and $i+1$ on route $r$ |
| $t_{r,i,i+1}$ | Driving time between consecutive locations $i$ and $i+1$ on route $r$ |
| $e_{r,i,i+1}$ | Energy consumption between consecutive locations $i$ and $i+1$ on route $r$ |
| $\Gamma^{\max}$ | Cap on the number of parallel sub-flows a route may be split into |
| $q$ | Sub-flow $q$ carries a proportion of the route $r$'s peak-hour flow |
| Energy parameters: | |
| $e_b$ | Energy capacity of each battery |
| $\lambda_{rib,p}^{arrive}$, $\lambda_{rib,p}^{depart}$ | (nonnegative real, SOS2) interpolation weight on breakpoint p of the charge curve, for battery *b*'s arrival/departure SoC at location $i$ on route $r$ |
| $\Phi_{rib}^{arrive}, \Phi_{rib}^{depart}$ | (nonnegative real) piecewise- linear interpolated time to charge from 0% SoC to battery *b*'s arrival/departure SoC at location $i$ on route $r$ |
| $s_p$ | SoC value at breakpoint $p = 0, \ldots, K$, with $s_0 = 0$ and $s_K = 1$ from the charge-curve data |
| $\varphi_p$ | Time (minutes) required to charge a battery from 0% SoC to $s_p$ , with $\varphi_0 = 0$ |
| $\rho^{charge}$ | Charger throughput (trucks per hour per charger), calculated as $\rho^{charge} = 60/\varphi_p$ |
| $\rho^{swap}$ | Swap throughput per hour per battery slot, calculated as $\rho^{swap} = 60/t^{swap}$ |
| $s_{min}$ | SoC floor at every arrival/departure, to avoid range anxiety (e.g. 10%) |
| $s_{charge}^{max}$ | Ceiling on SoC reachable via a charger (e.g. 100% for a full charge, or 80% for faster charging and battery health) |
| Station parameters: | |
| $C_i$ | Fixed upfront cost of building station $i$ |
| $\iota$ | Lifetime of station |
| $\alpha$ | Annual discount rate used in the capital recovery factor |
| $\rho$ | Capital recovery factor $\alpha(1+\alpha)^{\iota}/((1+\alpha)^{\iota}-1)$ |
| $N_i^{max}$ | Maximum installable chargers at candidate station $i$ (upper bounds on $N_i$) |
| $H_i^{max}$ | Maximum batteries at candidate station $i$ (upper bounds on $H_i$) |
| Time-related parameters: | |
| $t^{swap}$ | Time required to swap a battery |
| $t^{break}$ | Minimum short break time for consecutive driving |
| $t^{full}$ | Time required to charge a battery from 0% to 100% SoC at |
| $t^{rest}$ | Minimum long rest time for cumulative driving |
| $t^{drive}$ | Maximum consecutive driving time before having a $t^{break}$ break |
| $t^{trip}$ | Maximum cumulative driving time before having a $t^{rest}$ rest |
| days | Total operational days in a year (assumed 365 days) |
| VoT | Driver value of time (assumed 3 AUD per truck-minute) |
| Other parameters: | |
| $M$ | A sufficiently large positive constant (big-*M*) |
| $\varepsilon$ | A sufficiently small positive constant |
| Strategic (infrastructure) decision variables: | |
| $X_i$ | (binary) 1 if a station is deployed in location $i$, 0 otherwise |
| $Z_{rib}^{charge}$ | (binary) 1 if battery $b$ is charged at location $i$ on route $r$, 0 otherwise |
| $Z_{rib}^{swap}$ | (binary) 1 if battery $b$ is swapped at location $i$ on route $r$, 0 otherwise |
| $\omega_{riq}$ | (binary) 1 if sub-flow $q$ of route $r$ charges at location $i$ occupying one charger bay, 0 otherwise |
| $N_i$ | (nonnegative integer) number of deployed chargers at station $i$ at each demand stage |
| $H_i$ | (nonnegative integer) number of deployed full batteries in station $i$ at each demand stage |
| $\aleph_{rib,p}^{*}$ | (binary) Local linearisation auxiliary variable; if equal 1, it is an active segment for battery's SoC interpolation at location $i$ on route $r$ |
| Operational (scheduling) variables: | |
| $S_{rib}^{arrive}$ | (nonnegative real) SoC of battery $b$ upon arrival at location $i$ on route $r$ |
| $S_{rib}^{depart}$ | (nonnegative real) SoC of battery $b$ when it departs from visit $i$ on route $r$ |
| $T_{ri}^{arrive}$ | (nonnegative real) arrival time at location $i$ on route $r$ |
| $T_{ri}^{depart}$ | (nonnegative real) departure time from location $i$ on route $r$ |
| $T_{rib}^{charge}$ | (nonnegative real) charging time of battery $b$ at location $i$ on route $r$ |
| $D_{ri}$ | (nonnegative real) driver's cumulative working time (including short breaks and charging/swapping time) since last long rest, when arriving at location $i$ on route $r$ |
| $G_{ri}$ | (nonnegative real) driver's consecutive driving time since last long rest or break, when the driver arrives at location $i$ on route $r$ |
| $U_{ri}^{rest}$ | (nonnegative real) long rest duration at location $i$ on route $r$ |
| $U_{ri}^{break}$ | (nonnegative real) short break duration at location $i$ on route $r$ |
| $V_{ri}^{rest}$ | (binary) 1 if a $t^{rest}$ long rest occurs at location $i$ on route $r$, 0 otherwise |
| $V_{ri}^{break}$ | (binary) 1 if a $t^{break}$ short break occurs at location $i$ on route $r$, 0 otherwise |
| $\tau_{ri}, \eta_{ri}$ | Auxiliary variables for linearising delay objective function for charge/swap at location $i$ on route $r$, respectively |
| Multistage stochastic (SDDiP) notation: | |
| $n$ | Number of candidate stations on the route, $n = \lvert I_r \rvert$ |
| $\xi_t \in \Xi$ | Demand state in stage $t$ (Low, Mid or High adoption) |
| $P = [P_{jk}]$ | Transition matrix of the demand-state Markov chain ($j$: current state, $k$: next state) |
| $\kappa(\xi)$ | Demand multiplier of demand state $\xi$ (0.7, 1.0 and 1.3 for Low, Mid and High) |
| $\delta$ | Per-stage discount factor (0.94) |
| $x_t = (x_{t,i})_{i \in I_r}$ | (binary vector) station-deployment state at the end of stage *t*, with $x_{t,i} = X_i$ in stage *t* |
| $z_t$ | Local copy of the incoming state $x_{t-1}$ in stage *t* |
| $y_t$, $\mathcal{Y}_t(x_t, \xi_t)$ | Local (operational) decisions of stage *t* and their feasible set |
| $c_t$, $\tilde{c}_t$, $\Delta_t$ | Stage cost; operating (delay) cost; total daily truck delay (minutes) |
| $Q_t(x_{t-1}, \xi_t)$ | Value function: optimal expected discounted cost from stage *t* onward |
| $\mathcal{Q}_{t+1}(x, j)$, $\hat{\mathcal{Q}}_{t+1}(x, j)$ | Expected cost-to-go after stage *t* given deployment state *x* and current demand state *j*, and its cut approximation |
| $\mathcal{C}_{t,j}$, $\gamma_\ell$, $\beta_\ell$, $\psi$ | Cut set for stage *t* and demand state *j*; intercept and slope of cut ℓ; epigraph variable of the cost-to-go |
| $\hat{x}$ | Trial state at which cuts are generated |
| $\pi$, $\Lambda_{t,k}(\pi)$, $\underline{\Lambda}_{t,k}(\pi)$, $\mathcal{L}_{t,k}(\pi; \hat{x})$ | (non-negative) Lagrangian multipliers on the copy constraint, bounded by $\rho C_i$ (Proposition 2); dual-function constant; any lower bound on it (e.g. the solver's proven bound); Lagrangian dual function |
| $\mu_t$ | Policy (decision rule) of stage *t* |

The first objective term minimises annualised station setup cost, where the upfront establishment cost is converted to an equivalent annual cost over the planning horizon using the capital recovery factor, which converts the initial capital investment into an equivalent uniform annual cost, as follows:

$$\min \quad Z_1 = \sum_{i \in I} C_i X_i \left(\frac{\alpha(1+\alpha)^\iota}{(1+\alpha)^\iota - 1}\right) \tag{1}$$

where $C_i$ is the upfront establishment cost of stations, $\alpha$ is the annual discount rate, and $\iota$ is the lifetime of a station.

The second objective minimises the total annual delay cost incurred by trucks, calculated as the product of each truck's delay, the driver value of time (VoT), and the total operational days in a year (days). The delay at each station is determined by the time required for battery charging or swapping and, where applicable, mandatory rest and break requirements under HoS regulations. For swapping batteries, the fixed 10-minute swap time is independent of how many batteries are swapped. The second objective term is:

$$\min \quad Z_2 = \text{VoT} \times \text{days} \times \sum_{r \in R} f_r \sum_{i \in I_r} \left(T_{ri}^{depart} - T_{ri}^{arrive}\right)$$

$$= \text{VoT} \times \text{days} \times \sum_{r \in R} f_r \sum_{i \in I_r} \max\left(\max_b T_{rib}^{charge},\ t^{swap}\left\lceil \frac{1}{M} \sum_b Z_{rib}^{swap} \right\rceil,\ U_{ri}^{rest},\ U_{ri}^{break}\right) \tag{2}$$

To linearise this delay objective function term, we introduce new variables $\tau_{ri}$ and $\eta_{ri}$, where $\tau_{ri} \geq 0$ and $\eta_{ri} \in \{0,1\}$. Therefore, the linearised objective function of total delay can be replaced with:

$$\min \quad Z_2 = \text{VoT} \times n_{\text{day}} \times \sum_{r \in R} f_r \sum_{i \in I_r} \left(T_{ri}^{depart} - T_{ri}^{arrive}\right) \tag{3}$$

s.t.

$$\tau_{ri} \geq T_{rib}^{charge} \qquad \forall\, b \in B, i \in I_r, r \in R \tag{4}$$

$$\eta_{ri} \geq \frac{1}{M} \sum_b Z_{rib}^{swap} \qquad \forall\, i \in I_r, r \in R \tag{5}$$

$$T_{ri}^{depart} - T_{ri}^{arrive} \geq \tau_{ri} \qquad \forall\, i \in I_r, r \in R \tag{6}$$

$$T_{ri}^{depart} - T_{ri}^{arrive} \geq t^{swap}\eta_{ri} \qquad \forall\, i \in I_r, r \in R \tag{7}$$

$$T_{ri}^{depart} - T_{ri}^{arrive} \geq U_{ri}^{rest} \qquad \forall\, i \in I_r, r \in R \tag{8}$$

$$T_{ri}^{depart} - T_{ri}^{arrive} \geq U_{ri}^{break} \qquad \forall\, i \in I_r, r \in R \tag{9}$$

The formulation above allows every station location to choose freely between charging and swapping. However, motivated by the existing swapping business models around the world, the experimental study instead evaluates two separate, specialised variants, obtained by fixing the corresponding decision variables to zero and removing the associated capacity and cost terms. Furthermore, because swapping is typically far faster than a full charge ($\rho^{swap} \gg \rho^{charge}$, e.g. a battery swap in 10 minutes versus a full charge of more than an hour), separating two decisions will yield a clearer picture about each charge/swap decision. Accordingly, for a Charge model, no battery-swapping infrastructure is deployed ($Z^{swap} \equiv 0$), and for a Swap model ($Z^{charge} \equiv 0$), no charging infrastructure is deployed. Both are special cases of the general model above; the facility-capacity, energy, and HoS constraints are otherwise unchanged. The general delay objective's linearisation collapses accordingly, since the now-fixed decision variable makes one of $\tau_{ri}$ or $\eta_{ri}$ identically zero. The results reported in Section 5 are obtained under each specialisation, where the charge-only and swap-only specialisations use only the corresponding one of the two ceiling terms below:

$$T_{ri}^{depart} - T_{ri}^{arrive} \geq max\left(\max_b T_{rib}^{charge}, U_{ri}^{rest}, U_{ri}^{break}\right) \quad \forall i \in I_r, r \in R \tag{10}$$

$$T_{ri}^{depart} - T_{ri}^{arrive} \geq max\left(t^{swap} \cdot \eta_{ri}, U_{ri}^{rest}, U_{ri}^{break}\right) \quad \forall i \in I_r, r \in R \tag{11}$$

The following constraints ensure that replenishment can occur only at a deployed facility.

$$Z_{rib}^{charge} \leq X_i \quad , \quad Z_{rib}^{swap} \leq X_i \qquad \forall\, b \in B, r \in R,\ i \in I_r \tag{12}$$

If a battery module is not selected for charging at a visit, its charging time is zero at that visit.

$$T_{rib}^{charge} \leq M Z_{rib}^{charge} \qquad \forall\, b \in B, r \in R, i \in I_r \tag{13}$$

A positive charging time activates the corresponding charging decision, subject to the numerical tolerance used in the formulation.

$$Z_{rib}^{charge} \leq M T_{rib}^{charge} \qquad \forall\, b \in B, r \in R, i \in I_r \tag{14}$$

A battery module cannot be charged and exchanged during the same visit.

$$Z_{rib}^{charge} + Z_{rib}^{swap} \leq 1 \qquad \forall\, b \in B, r \in R, i \in I_r \tag{15}$$

If a battery module is neither charged nor exchanged at a facility, its SoC remains unchanged during the visit.

$$S_{rib}^{depart} - S_{rib}^{arrive} \leq Z_{rib}^{swap} + Z_{rib}^{charge} \qquad \forall\, b \in B, r \in R, i \in I_r \tag{16}$$

For a charged battery module, departure SoC cannot be lower than arrival SoC.

$$S_{rib}^{depart} \geq S_{rib}^{arrive} \qquad \forall\, b \in B, r \in R, i \in I_r \tag{17}$$

An exchanged battery module is assigned the assumed replacement SoC of 100%.

$$S_{rib}^{depart} \geq Z_{rib}^{swap} \qquad \forall\, b \in B, r \in R, i \in I_r \tag{18}$$

The SoC balance accounts for depletion during travel and replenishment during a charging or swapping event.

$$S_{r,i+1,b}^{arrive} = S_{rib}^{depart} - \frac{e_{r,i,i+1}}{e_b \cdot |B|} \;\; \forall\, b \in B, i \in I_r \backslash \{d_r\}, r \in R \tag{19}$$

The initial SoC of every battery module at the origin is fixed at 100% in the numerical experiments.

$$S_{r,O_r,b}^{depart} = 1 \;\; \forall\, b \in B, r \in R \tag{20}$$

To provide a minimum range reserve, the arrival SoC at each facility visit is constrained to be no lower than the specified minimum SoC.

$$S_{rib}^{arrive} \geq s_{\min}, \quad S_{rib}^{depart} \geq s_{\min} \quad \forall\, b \in B,\ i \in I_r,\ r \in R \tag{21}$$

If a battery is charged at a station, its SoC cannot exceed the maximum SoC ($s_{charge}^{max}$). This parameter can be set to 100% similar to swapped new batteries, or be considered as 80% for minimising the charging delay and/or for battery health reasons.

$$S_{rib}^{depart} \leq s_{charge}^{max} + \left(1 - s_{charge}^{max}\right)\left(1 - Z_{rib}^{charge}\right) \;\; \forall\, b \in B, i \in I_r, r \in R \tag{22}$$

This study assumes that battery charging follows a constant-current/constant-voltage (CC-CV) profile, in which the charge rate is roughly constant up to about 80% SoC, then tapers as SoC approaches 100%. Equivalently, the time required to reach a given SoC, starting from empty, is a convex, increasing function of SoC where equal increments of SoC take longer to add as SoC rises. Hence, at every location, the battery's arrival and departure SoC are each written as a convex combination of the *K+1* charge-curve breakpoints:

$$\lambda_{rib,p}^{arrive}, \lambda_{rib,p}^{depart} \geq 0, \quad \sum_{p=0}^{K} \lambda_{rib,p}^{arrive} = \sum_{p=0}^{K} \lambda_{rib,p}^{depart} = 1 \quad \forall\, b \in B,\ i \in I_r,\ r \in R \tag{23}$$

$$S_{rib}^{arrive} = \sum_{p=0}^{K} \lambda_{rib,p}^{arrive}\, s_p, \quad S_{rib}^{depart} = \sum_{p=0}^{K} \lambda_{rib,p}^{depart}\, s_p \quad \forall\, b \in B,\ i \in I_r,\ r \in R \tag{24}$$

$$\Phi_{rib}^{arrive} = \sum_{p=0}^{K} \lambda_{rib,p}^{arrive}\, \varphi_p, \quad \Phi_{rib}^{depart} = \sum_{p=0}^{K} \lambda_{rib,p}^{depart}\, \varphi_p \quad \forall\, b \in B,\ i \in I_r,\ r \in R \tag{25}$$

where $s_p$ represents the SoC value at the breakpoint $p = 0, \ldots, K$, with $s_0 = 0$ and $s_K = 1$ from the charge-curve data. $\varphi_p$ is the time required to charge a battery from 0% SoC to $s_p$ , with $\varphi_0 = 0$. Since the curve is convex, the implied slopes $(\varphi_p - \varphi_{p-1})/(s_p - s_{p-1})$ are increasing in $p$. The set $\{\lambda^*_{rib,0}, \ldots, \lambda^*_{rib,K}\}$ is an SOS2 set (at most two consecutive nonzero), for all $b \in B$, $r \in R$, $i \in I_r$, where * denotes either arrive or depart. Because $\{\lambda^*_{rib,p}\}$ is an SOS2 set, $\Phi^*_{rib}$ is a piecewise-linear interpolation of the charging time curve equivalent at $S^*_{rib}$. Accordingly, at most two, necessarily consecutive, weights are nonzero. This makes the interpolation exact and continuous over the whole [0,1] SoC range.

The physically required charging time is then pinned to the difference between the interpolated cumulative-charge-time values at departure and arrival, activated only when the battery is actually charged (big-M linearisation, when $Z^{charge}_{rib}$=0). In other words, if a battery is charged in a location ($Z^{charge}_{rib} = 1$), the charging time must equal the (piecewise-linearised) physically required time for that SoC gain:

$$T^{charge}_{rib} \leq \Phi^{depart}_{rib} - \Phi^{arrive}_{rib} + M\left(1 - Z^{charge}_{rib}\right) \ \forall\, b \in B, i \in I_r, r \in R \tag{26}$$

$$T^{charge}_{rib} \geq \Phi^{depart}_{rib} - \Phi^{arrive}_{rib} - M\left(1 - Z^{charge}_{rib}\right) \ \forall\, b \in B, i \in I_r, r \in R \tag{27}$$

**Remark 1:** When $Z^{charge}_{rib} = 1$, the two constraints sandwich $T^{charge}_{rib}$ to equal exactly $\Phi^{depart}_{rib} - \Phi^{arrive}_{rib}$ – the piecewise-linearised physical charging time for that SoC transition.

**Remark 2:** When $Z^{charge}_{rib} = 0$, $T^{charge}_{rib} \leq MZ^{charge}_{rib}$ forces $T^{charge}_{rib} = 0$, and the "SoC unchanged if not charged" constraint forces $S^{depart}_{rib} = S^{arrive}_{rib}$, hence $\lambda^{depart}_{rib} = \lambda^{arrive}_{rib}$ (same weights) and $\Phi^{depart}_{rib} = \Phi^{arrive}_{rib}$ automatically.

**Remark 3:** Because $\{\lambda^*_{rib,p}\}$ is an SOS2 set, $\Phi^*_{rib}$ is pinned exactly to the piecewise-linear value of the charging curve at $S^*_{rib}$ – in both directions – regardless of how $\Phi^*_{rib}$ enters the objective or other constraints. There is no scope for the optimiser to inflate or deflate $\Phi^{arrive}_{rib}$ to artificially relax the charging-time bound.

**Remark 4:** SOS2 constraints are natively supported by standard MILP solvers (e.g. Gurobi and CPLEX) and the solver's branching handles the "at most two consecutive nonzero" logic internally.

Each route's schedule may use at most $N_i$ chargers / $H_i$ batteries at station $i$. Since the peak-hour demand ($f^{peak}_r$) is significantly higher than the station capacities, we introduce a route-splitting method so a single route's demand can be met by more than one station when pooled capacity requires it. This demand splitting is needed because with a single schedule per route, a route's entire concurrent demand must be served by whichever one station its schedule visits – if that demand exceeds even the largest station's throughput, the model is infeasible, even when spreading the same demand across two or three nearby stations would be feasible. Route-splitting resolves this by representing each route as several parallel sub-flows, each free to choose its own station. The number of parallel sub-flows $|\Gamma_r|$ a route is split into is the smallest number such that the route's peak-hour flow, divided evenly across sub-flows, does not exceed the throughput of the bottleneck candidate station (most capacity-constrained) along the route.

**Remark 5:** Every decision variable and constraint previously indexed by battery $b$ alone is now indexed by the pair $(q, b)$, $q \in \Gamma_r$, $b \in B$. Therefore, the notations $S^{arrive}_{rib}$, $S^{depart}_{rib}$, $T^{charge}_{rib}$, $Z^{charge}_{rib}$, $Z^{swap}_{rib}$, $\lambda^*_{rib,p}$, $\Phi^*_{rib}$ all become $S^{arrive}_{riqb}$, $S^{depart}_{riqb}$, $T^{charge}_{riqb}$, $Z^{charge}_{riqb}$, $Z^{swap}_{riqb}$, $\lambda^*_{riqb,p}$, $\Phi^*_{riqb}$ respectively, with $q \in \Gamma_r$ added to every $\forall$ quantifier that previously read $b \in B$. The route-level timeline variables $T^{arrive}_{ri}$, $T^{depart}_{ri}$, $D_{ri}$, $G_{ri}$, $U^{rest}_{ri}$, $U^{break}_{ri}$, $V^{rest}_{ri}$, $V^{break}_{ri}$ are not widened, because every sub-flow of a route shares the same arrival/departure schedule, since dwell time depends on whether/how long a stop takes, not on which specific station handles it. Sharing one timetable means that a sub-flow that does not charge at a stop still dwells there for as long as the sub-flow that does, which is conservative for delay.

Charging capacity constraints limit the concurrent peak demand that can be served by the installed charging units. After route demand is divided across sub-flows, charger capacity is counted per truck rather than per battery, because a whole truck occupies one charger bay while all its battery modules charge in parallel. A charging-occupancy indicator $\omega_{riq}$ is therefore introduced for each visit and sub-flow, equal to 1 if sub-flow $q$ of route $r$ charges at location $i$, and 0 otherwise. It is linked to the battery-level charging decisions and to the installed chargers by:

$$Z_{riqb}^{charge} \leq \omega_{riq} \quad \forall\, b \in B,\; q \in \Gamma_r,\; i \in I_r,\; r \in R\,, \quad \text{and} \quad \sum_{q\in\Gamma_r} \omega_{riq} \leq N_i \quad \forall\, i \in I_r,\; r \in R \tag{28}$$

The pooled peak-hour throughput constraint is then:

$$\sum_{r\in R} \sum_{q\in\Gamma_r} \frac{f_r^{peak}}{|\Gamma_r|}\, \omega_{riq} \leq \rho^{charge} N_i, \quad \rho^{charge} = \frac{90}{t^{full}} \quad \forall\, i \in I \tag{29}$$

where $f_r^{peak}$ is the BET peak-hour flow of route $r$ in the current stage (trucks per hour) and $t^{full}$ is the time to charge a battery from 0% to 100% SoC (90 minutes in the experiments). Each charger can therefore serve $90/t^{full}$ trucks per hour, conservatively assuming that every charging event occupies the charger for a full-charge time. The swap model has the analogous constraint, counted per battery because $H_i$ is a battery inventory:

$$\sum_{r\in R} \sum_{q\in\Gamma_r} \sum_{b\in B} \frac{f_r^{peak}}{|\Gamma_r|}\, Z_{riqb}^{swap} \leq \rho^{swap} H_i, \text{and} \quad \rho^{swap} = \frac{60}{t^{swap}} \quad \forall\, i \in I \tag{30}$$

The delay-linearisation constraints are widened to include the sub-flow index:

$$\tau_{ri} \geq T_{riqb}^{charge} \quad \forall\, q \in \Gamma_r,\; b \in B,\; i \in I_r,\; r \in R \tag{31}$$

$$\eta_{ri} \geq \frac{1}{M} \sum_{q\in\Gamma_r} \sum_{b\in B} Z_{riqb}^{swap} \quad \forall\, i \in I_r,\; r \in R \tag{32}$$

Each sub-flow represents an independent group of trucks with its own batteries, not additional batteries on the same truck, so the energy-balance and battery-synchronisation constraints hold within each sub-flow separately, not summed across sub-flows:

$$S_{r,i+1,q,b_1}^{arrive} - S_{riqb_1}^{depart} = S_{r,i+1,q,b_2}^{arrive} - S_{riqb_2}^{depart} \quad \forall\, b_1 \neq b_2 \in B,\; q \in \Gamma_r,\; r \in R,\; i \in I_r \backslash \{d_r\} \tag{33}$$

$$\sum_{b\in B} S_{riqb}^{depart} - \sum_{b\in B} S_{r,i+1,q,b}^{arrive} = \frac{e_{r,i,i+1}}{e_b} \quad \forall\, q \in \Gamma_r,\; r \in R,\; i \in I_r \backslash \{d_r\} \tag{34}$$

Remark 6: These are the only constraints that change in substance; every other constraint (station availability, charge/swap mutual exclusivity, HoS rules, travel-time propagation, and the charging-curve block) carries over unchanged under the mechanical $(q, b)$ index-widening described above.

Remark 7: When $|\Gamma_r| = 1$ for every route (e.g. $\Gamma^{\max} = 1$, or no route's peak flow requires splitting), every sub-flow set is a singleton and the model above reduces to the formulation without route splitting: $\omega_{riq}$ becomes a single per-visit charging indicator, and Eqs. (29) and (30) apply to the full peak-hour flow $f_r^{peak}$ of each route.

Remark 8: The reformulation multiplies every per-battery variable and constraint on route $r$ by $|\Gamma_r|$; total model size scales with $\sum_{r\in R}|\Gamma_r|$ rather than $|R|$. Since $|\Gamma_r|$ is capped at $\Gamma^{\max}$, a route requiring heavy splitting (busy corridor, tight station capacity) costs proportionally more to solve.

Rest break regulatory constraints: A driver cannot take both a long rest and short break at the same visit.

$$V_{ri}^{break} + V_{ri}^{rest} \le 1 \qquad \forall\, r \in R, i \in I_r \tag{35}$$

A long rest can be taken only at a deployed station, $V_{ri}^{rest} \le X_i$, as for charging and swapping. The drivers' long rest constraints are formulated as follows:

$$U_{ri}^{rest} \ge t^{rest} V_{ri}^{rest} \quad, U_{ri}^{rest} \le M V_{ri}^{rest} \quad, 0 \le D_{ri} \le t^{trip} \qquad \forall\, r \in R, i \in I_r \tag{36}$$

The drivers' short break constraint is formulated as follows:

$$U_{ri}^{break} \ge t^{break} V_{ri}^{break} \quad, U_{ri}^{break} \le M V_{ri}^{break} \quad, 0 \le G_{ri} \le t^{drive} \qquad \forall\, r \in R, i \in I_r \tag{37}$$

Travel time propagation is formulated as follows:

$$T_{r,i+1}^{arrive} = T_{ri}^{depart} + t_{r,i,i+1} \qquad \forall\, r \in R, i \in I_r \backslash \{d_r\} \tag{38}$$

If a driver has a long rest at a location $i_1$ (i.e., $V_{r,i_1}^{rest} = 1$), and does not have long rests in locations $i_1 + 1, \dots, i_2 - 1$ (i.e., $\sum_{i=i_1+1}^{i_2-1} V_{ri}^{rest} = 0$), then the cumulative working time when the driver arrives at $i_2$ is $(T_{r,i_2}^{arrive} - T_{r,i_1}^{depart})$ (i.e., $D_{r,i_2} = T_{r,i_2}^{arrive} - T_{r,i_1}^{depart}$). This is formulated as:

$$D_{r,i_2} - \left(T_{r,i_2}^{arrive} - T_{r,i_1}^{depart}\right) \le M\left(1 - \sum_{i=i_1+1}^{i_2-1} V_{ri}^{rest} - \varepsilon\right) \qquad \forall\, r \in R, i_1 < i_2 \in I_r \tag{39}$$

$$D_{r,i_2} - \left(T_{r,i_2}^{arrive} - T_{r,i_1}^{depart}\right) \ge \varepsilon\left(1 - V_{r,i_1}^{rest}\right) - M \sum_{i=i_1+1}^{i_2-1} V_{ri}^{rest} \qquad \forall\, r \in R, i_1 < i_2 \in I_r \tag{40}$$

$$D_{r,i_2} - \left(T_{r,i_2}^{arrive} - T_{r,i_1}^{depart}\right) \le M\left(1 - V_{r,i_1}^{rest} + \sum_{i=i_1+1}^{i_2-1} V_{ri}^{rest}\right) \qquad \forall\, r \in R, i_1 < i_2 \in I_r \tag{41}$$

$$D_{r,i_2} - \left(T_{r,i_2}^{arrive} - T_{r,i_1}^{depart}\right) \ge -M\left(1 - V_{r,i_1}^{rest} + \sum_{i=i_1+1}^{i_2-1} V_{ri}^{rest}\right) \qquad \forall\, r \in R, i_1 < i_2 \in I_r \tag{42}$$

Similarly, if a driver has either a short break or long rest in location $i_1$ (i.e., $V_{r,i_1}^{break} + V_{r,i_1}^{rest} = 1$), and does not have any short breaks or long rests in locations $i_1 + 1, \dots, i_2 - 1$ (i.e., $\sum_{i=i_1+1}^{i_2-1}(V_{ri}^{break} + V_{ri}^{rest}) = 0$), then the consecutive driving time when the driver arrives at $i_2$ is $(T_{r,i_2}^{arrive} - T_{r,i_1}^{depart})$ (i.e., $G_{r,i_2} = T_{r,i_2}^{arrive} - T_{r,i_1}^{depart}$). This is formulated as:

$$G_{r,i_2} - \left(T_{r,i_2}^{arrive} - T_{r,i_1}^{depart}\right) \le M\left(1 - \sum_{i=i_1+1}^{i_2-1}\left(V_{ri}^{break} + V_{ri}^{rest}\right) - \varepsilon\right) \quad \forall\, r \in R, i_1 < i_2 \in I_r \tag{43}$$

$$G_{r,i_2} - \left(T_{r,i_2}^{arrive} - T_{r,i_1}^{depart}\right) \ge \varepsilon\left(1 - V_{r,i_1}^{break} - V_{r,i_1}^{rest}\right) - M \sum_{i=i_1+1}^{i_2-1}(V_{ri}^{break} + V_{ri}^{rest}) \, \forall\, r \in R, i_1 < i_2 \in I_r \tag{44}$$

$$G_{r,i_2} - \left(T_{r,i_2}^{arrive} - T_{r,i_1}^{depart}\right) \leq M\left(1 - V_{r,i_1}^{break} - V_{r,i_1}^{rest} + \sum_{i=i_1+1}^{i_2-1}(V_{ri}^{break} + V_{ri}^{rest})\right) \quad \forall\, r \in R, i_1 < i_2 \in I_r \quad (45)$$

$$G_{r,i_2} - \left(T_{r,i_2}^{arrive} - T_{r,i_1}^{depart}\right) \geq -M\left(1 - V_{r,i_1}^{break} - V_{r,i_1}^{rest} + \sum_{i=i_1+1}^{i_2-1}(V_{ri}^{break} + V_{ri}^{rest})\right) \quad \forall\, r \in R, i_1 < i_2 \in I_r \quad (46)$$

If the difference between the arrival and departure time at location $i$ is longer than or equal to $t^{rest}$, then the driver has a long rest at the location $i$ ($V_{ri}^{rest} = 1$). Otherwise, $V_{ri}^{rest} = 0$.

$$MV_{ri}^{rest} \geq T_{ri}^{depart} - T_{ri}^{arrive} - t^{rest} + \varepsilon \qquad \forall\, r \in R, i \in I_r \quad (47)$$

$$V_{ri}^{rest} - 1 \leq \varepsilon(T_{ri}^{depart} - T_{ri}^{arrive} - t^{rest}) \quad \forall\, r \in R, i \in I_r \quad (48)$$

If the difference between the arrival and departure time at location $i$ is longer than or equal to $t^{break}$ but shorter than $t^{rest}$, then the driver has a short break in this location $i$ ($V_{ri}^{break} = 1$). Otherwise $V_{ri}^{break} = 0$.

$$M(V_{ri}^{break} + V_{ri}^{rest}) \geq T_{ri}^{depart} - T_{ri}^{arrive} - t^{break} + \varepsilon \qquad \forall\, r \in R, i \in I_r \quad (49)$$

$$V_{ri}^{break} - 1 \leq \varepsilon(T_{ri}^{depart} - T_{ri}^{arrive} - t^{break}) \qquad \forall\, r \in R, i \in I_r \quad (50)$$

Remark 9: When $T_{ri}^{depart} - T_{ri}^{arrive} \geq t^{rest}$, constraints $MV_{ri}^{rest} \geq T_{ri}^{depart} - T_{ri}^{arrive} - t^{rest} + \varepsilon$ and $V_{ri}^{break} + V_{ri}^{rest} \leq 1$ force $V_{ri}^{rest} = 1$ and $V_{ri}^{break} = 0$.

Remark 10: When $t^{break} \leq T_{ri}^{depart} - T_{ri}^{arrive} < t^{rest}$, constraints $M(V_{ri}^{break} + V_{ri}^{rest}) \geq T_{ri}^{depart} - T_{ri}^{arrive} - t^{break} + \varepsilon$ force $V_{ri}^{break} + V_{ri}^{rest} = 1$, and constraints $V_{ri}^{rest} - 1 \leq \varepsilon(T_{ri}^{depart} - T_{ri}^{arrive} - t^{rest})$ make $V_{ri}^{rest} = 0$, so $V_{ri}^{break} = 1$.

Variable domains and linking conditions are as follows:

$$Z^{charge}, Z^{swap}, V^{rest}, V^{break} \in \{0,1\} \quad (51)$$

$$0 \leq S^{arrive} \leq 1, 0 \leq S^{depart} \leq 1 \quad (52)$$

$$T^{charge}, T^{arrive}, T^{depart} \geq 0 \quad (53)$$

$$0 \leq D \leq t^{trip},\ \ 0 \leq G \leq t^{drive} \quad (54)$$

$$U_{ri}^{rest} \geq 0, \quad U_{ri}^{break} \geq 0 \quad (55)$$

**Multistage stochastic formulation by SDDiP:** The SDDiP model is solved independently for each route $r$. Let $I_r$ be the set of candidate stations on the route, with $n = |I_r|$. Stages $t = 1, \ldots, T$ ($T = 4$) are the planning stages. In each stage, demand is the base BET penetration of that stage scaled by a multiplier $\kappa(\xi_t)$, where $\xi_t \in \Xi = \{\text{Low}, \text{Mid}, \text{High}\}$ is the demand state and $\kappa(\text{Low}) = 0.7$, $\kappa(\text{Mid}) = 1$, $\kappa(\text{High}) = 1.3$.

The state at the end of stage $t$ is the vector of station-opening decisions is defined as:

$$x_t = \left(x_{t,i}\right)_{i\in I_r} \in \{0,1\}^n \tag{56}$$

where $x_0 = 0$ and $x_{t,i}$=1 if station $i$ is open at the end of stage *t*. The demand state $\xi_t$ is an exogenous information state, so the full dynamic-programming state entering stage $t$ is $(x_{t-1}, \xi_t)$. All other decisions of stage *t* are local (stage) variables, for example $y_t$ represents the number of chargers $N_{t,i}$ (or batteries $H_{t,i}$) at each station. Other stage variables include the assignment of driver sub-flows to stations, arrival and departure times, state of charge, charging or swapping decisions, and short-break and long-rest indicators. Charger and battery counts are re-optimised in every stage, subject to the station being open and to the stage capacity limit $N_{t,i} \leq N_{t,i}^{\max} x_{t,i}$; they are not part of the state. Only the decision to open a station is irreversible. Because charger and battery costs are not included in the objective function, the reported numbers of chargers and batteries should be read as the capacity used by the scheduling decisions rather than as cost-optimised sizing.

The state-transition equation is dictated by the assumption that opening a station is irreversible:

$$x_t \geq x_{t-1}, \quad x_t \in \{0,1\}^n, \quad t = 1, \dots, T \tag{57}$$

Following Zou et al. (2019), a local copy $z_t$ of the incoming state is introduced:

$$z_t = x_{t-1}, \quad 0 \leq z_t \leq x_t, \quad z_t \in [0,1]^n \tag{58}$$

The copy constraint $z_t = x_{t-1}$ is the only link between consecutive stages, and it is the constraint that is dualised in the backward pass.

The demand state follows a first-order Markov chain with transition matrix $P = [P_{jk}]$, where $P_{jk} = \Pr(\xi_t = k \mid \xi_{t-1} = j)$, and $\xi_1 = \text{Mid}$ is known at the outset (the transition probabilities are given in Table 6, Section 5).

Within each stage, the sequence of events is: (i) the stage demand state $\xi_t$ is observed; (ii) the station-opening decisions $x_t \geq x_{t-1}$ are made, and new stations are available immediately; (iii) the operational decisions $y_t$ are made for the realised demand, and the stage cost is incurred; (iv) the process moves to stage *t* + 1, where $\xi_{t+1}$ is drawn from row $\xi_t$ of $P$. Decisions in stage $t$ are therefore made with full knowledge of the current demand but only the probability distribution of future demand.

Let $\xi_{[t]} = (\xi_1, \dots, \xi_t)$ denote the demand history. A policy is nonanticipative if $x_t$ and $y_t$ are functions of $\xi_{[t]}$ only; equivalently, in the extensive form on the scenario tree, two scenarios that share the same history up to stage $t$ must have identical stage-*t* decisions. Because the demand process is Markov and all inter-stage coupling passes through $x_{t-1}$, it is sufficient to consider Markov policies $x_t = \mu_t(x_{t-1}, \xi_t)$. The SDDiP policy in the stage-*t* problem uses only $x_{t-1}$, the observed $\xi_t$ and an approximation of the expected future cost conditional on $\xi_t$. It never uses future realisations, so it is nonanticipative by construction.

The cost incurred in stage *t* is defined as:

$$c_t(x_{t-1}, x_t, y_t; \xi_t) = \rho \sum_{i\in I_r} C_i \left(x_{t,i} - x_{t-1,i}\right) + \text{VoT}\, n_{\text{day}}\, \Delta_t(y_t; \xi_t) \tag{59}$$

where $C_i$ is the fixed cost of station $i$, $\rho$ is the capital recovery factor (6%, 20 years), VoT is the value of time (AUD 3 per truck-minute), $n_{\text{day}} = 365$ operating days, and $\Delta_t$ is the total daily truck delay in minutes, which depends on $\xi_t$ through the truck flow. Only newly opened stations incur capital cost.

The value function is defined as follows:

$$Q_t(x_{t-1}, \xi_t) = \min_{x_t, y_t, z_t} \{c_t(z_t, x_t, y_t; \xi_t) + \delta \mathcal{Q}_{t+1}(x_t, \xi_t)\} \tag{60}$$

$$\text{s.t.} \;\; z_t = x_{t-1}, \;\; 0 \le z_t \le x_t, \;\; x_t \in \{0,1\}^n, \;\; y_t \in \mathcal{Y}_t(x_t, \xi_t)$$

where $\delta = 0.94$ is the per-stage discount factor and $\mathcal{Y}_t(x_t, \xi_t)$ is the operationally feasible set of the stage MILP.

The expected cost-to-go function is then defined as:

$$\mathcal{Q}_{t+1}(x_t, j) = \sum_{k \in \Xi} P_{jk} Q_{t+1}(x_t, k), \quad Q_{T+1} \equiv 0 \tag{61}$$

The optimal expected discounted cost is $Q_1(x_0, \text{Mid})$

For the cost-to-go approximation, the SDDiP replaces $\mathcal{Q}_{t+1}(\cdot, j)$ by a polyhedral under-approximation built from cuts, with one cut family for each demand state $j$:

$$\hat{\mathcal{Q}}_{t+1}(x, j) = \max\left\{0, \max_{\ell \in \mathcal{C}_{t+1,j}} (\gamma_\ell + \beta_\ell^\top x)\right\} \tag{62}$$

The constant 0 is a valid lower bound because all costs are non-negative.

The policy is the rule that, in stage $t$ and given $(x_{t-1}, \xi_t)$, solves the value-function problem with $\mathcal{Q}_{t+1}$ replaced by $\hat{\mathcal{Q}}_{t+1}$ and implements the resulting $(x_t, y_t)$. Policy evaluation estimates the expected discounted cost $E[\sum_t \delta^{t-1} c_t]$ of this policy by simulation (see 'Lower bound, upper bound' below). It is distinct from both the value function and its approximation. The following ordering holds:

$$\text{LB} = \hat{Q}_1 \;\le\; Q_1 \;\le\; \text{expected cost of the SDDiP policy} \tag{63}$$

where $\hat{Q}_1$ is the stage-1 optimal value using the cut approximation.

***Lagrangian cuts and conditions for validity:*** For a trial state $\hat{x} \in \{0,1\}^n$ and demand state $k$ in stage $t$, the copy constraint $z_t = \hat{x}$ is dualised with multiplier $\pi \in \mathbb{R}^n$:

$$\mathcal{L}_{t,k}(\pi; \hat{x}) = \min_{x,y,z} \left\{\tilde{c}_t(x, y; k) + \rho C^\top (x - z) + \delta \hat{\mathcal{Q}}_{t+1}(x, k) + \pi^\top (z - \hat{x})\right\} \tag{64}$$

$$\text{s.t.} \;\; 0 \le z \le x, \;\; x \in \{0,1\}^n, \;\; y \in \mathcal{Y}_t(x, k)$$

where $\tilde{c}_t = \text{VoT}\, n_{\text{day}}\, \Delta_t$ is the operating cost. The variable $z$ appears only in separable terms, so it can be minimised in closed form, $\min_{0 \le z_i \le x_i} (\pi_i - \rho C_i) z_i = \min(0, \pi_i - \rho C_i)\, x_i$. This gives

$$\mathcal{L}_{t,k}(\pi; \hat{x}) = \Lambda_{t,k}(\pi) - \pi^\top \hat{x}, \quad \Lambda_{t,k}(\pi) = \min_{x,y} \left\{\tilde{c}_t + \sum_i \min(\rho C_i, \pi_i)\, x_i + \delta \hat{\mathcal{Q}}_{t+1}(x, k)\right\} \tag{65}$$

where the dual function is evaluated by solving one MILP with modified station costs. The cut passed to stage $t$-1 for demand state $j$ is the probability-weighted aggregate of the child cuts:

$$\psi \ge \sum_k P_{jk} \left(\Lambda_{t,k}(\pi_k) - \pi_k^\top x\right) = \gamma + \beta^\top x, \quad \beta = -\sum_k P_{jk}\, \pi_k \tag{66}$$

***Proposition 1 (validity of the Lagrangian cuts):*** Let $\hat{x} \in \{0,1\}^n$ be a trial state in stage $t$, and for each demand state $k \in \Xi$ let $\pi_k \in \mathbb{R}^n$ be any multiplier vector. Let $\underline{\Lambda}_{t,k}(\pi_k)$ be any lower bound on the optimal value $\Lambda_{t,k}(\pi_k)$ of the MILP in Eq. (65), such as the solver's proven bound. For a parent demand state $j$, define the cut coefficients as follows:

$$\gamma = \sum_{k \in \Xi} P_{jk}\, \underline{\Lambda}_{t,k}(\pi_k), \qquad \beta = -\sum_{k \in \Xi} P_{jk}\, \pi_k \tag{67}$$

Assume Condition 1 where the cut approximation of the next stage satisfies $\hat{\mathcal{Q}}_{t+1}(\cdot, k) \le \mathcal{Q}_{t+1}(\cdot, k)$ for all $k$, and Condition 2 where the stage-$t$ problem is feasible for every incoming state, so that every $\Lambda_{t,k}$ is finite. Then

$$\gamma + \beta^\top x_{t-1} \le \mathcal{Q}_t(x_{t-1}, j) \quad \text{for every } x_{t-1} \in \{0,1\}^n \tag{68}$$

The cut is therefore a valid lower bound on the expected cost-to-go at every state $x_{t-1}$, not only at the trial state $\hat{x}$ where it was generated.

***Proof:*** Fix any $x_{t-1} \in \{0,1\}^n$ and any $k \in \Xi$. If $Q_t(x_{t-1}, k) = +\infty$ the inequality below holds trivially, so assume it is finite and let $(x^*, y^*)$ be an optimal solution of the stage-$t$ problem in Eq. (60) with incoming state $x_{t-1}$ and demand state $k$. Then $x^* \ge x_{t-1}$, $x^* \in \{0,1\}^n$ and $y^* \in \mathcal{Y}_t(x^*, k)$.

Consider the Lagrangian problem in Eq. (64) with the trial state $\hat{x}$ replaced by $x_{t-1}$, that is, with the copy constraint $z = x_{t-1}$ dualised. The point $(x, y, z) = (x^*, y^*, x_{t-1})$ is feasible for it, because $0 \le x_{t-1} \le x^*$. At this point the penalty term $\pi_k^\top(z - x_{t-1})$ is zero, and $\rho C^\top(x^* - x_{t-1})$ is exactly the capital cost of the stations opened in stage $t$. Hence:

$$\mathcal{L}_{t,k}(\pi_k; x_{t-1}) \le \tilde{c}_t(x^*, y^*; k) + \rho C^\top(x^* - x_{t-1}) + \delta\, \hat{\mathcal{Q}}_{t+1}(x^*, k) \le c_t(x_{t-1}, x^*, y^*; k) + \delta\, \mathcal{Q}_{t+1}(x^*, k) = Q_t(x_{t-1}, k) \tag{69}$$

where the second inequality uses the Condition1. By Eq. (65), $\mathcal{L}_{t,k}(\pi_k; x_{t-1}) = \Lambda_{t,k}(\pi_k) - \pi_k^\top x_{t-1}$, and $\underline{\Lambda}_{t,k} \le \Lambda_{t,k}$, so $\underline{\Lambda}_{t,k}(\pi_k) - \pi_k^\top x_{t-1} \le Q_t(x_{t-1}, k)$. Multiplying by $P_{jk} \ge 0$, summing over $k$ and using Eq. (61) gives

$$\gamma + \beta^\top x_{t-1} = \textstyle\sum_{k\in\Xi} P_{jk} \left(\underline{\Lambda}_{t,k}(\pi_k) - \pi_k^\top x_{t-1}\right) \le \sum_{k\in\Xi} P_{jk}\ Q_t(x_{t-1}, k) = \mathcal{Q}_t(x_{t-1}, j) \tag{70}$$

Finally, the Condition1 holds in the last stage because $\hat{\mathcal{Q}}_{T+1} = \mathcal{Q}_{T+1} \equiv 0$. If it holds at stage $t+1$, the argument above shows that every cut added to stage $t$ is valid. Since $\hat{\mathcal{Q}}_t$ is the maximum of 0 and valid cuts, and $\mathcal{Q}_t \ge 0$, it follows that $\hat{\mathcal{Q}}_t \le \mathcal{Q}_t$. By backward induction, the Condition1 holds at every stage, and so every cut generated by the algorithm is valid.

Remark 11: The trial state $\hat{x}$ influences the cut only through the choice of the multipliers $\pi_k$, which Algorithm 3 selects to make the cut as tight as possible at $\hat{x}$. Validity holds for any $\pi_k$ and any lower bound $\underline{\Lambda}_{t,k}$. Ending the level method early, or stopping a MILP at its time limit, therefore weakens a cut but never invalidates it. Tightness of the cut at $\hat{x}$, which Zou et al. (2019) (Theorem 3) establish for binary states, additionally requires that $\pi$ maximise the dual and that $\Lambda$ be evaluated exactly.

***Proposition 2 (Sign and bounds of the multipliers)*:** Suppose the operating cost $h_{t,k}(x) = \min_{y\in\mathcal{Y}_t(x,k)} \tilde{c}_t(x, y; k) + \delta\hat{\mathcal{Q}}_{t+1}(x, k)$ is non-increasing in $x$. Then

$$\sup_{\pi\in\mathbb{R}^n} \mathcal{L}_{t,k}(\pi; \hat{x}) = \sup_{\pi\in[0,\rho C]} \mathcal{L}_{t,k}(\pi; \hat{x}), \tag{71}$$

Therefore, the multipliers can be restricted to the box $0 \le \pi_i \le \rho C_i$ without loss.

Monotonicity of *h:* In the stage model, $x_i$ appears only in enabling constraints: $N_i \le N_i^{\max} x_i$, $H_i \le H_i^{\max} x_i$, and the charging, swapping and long-rest indicators at station $i$ are bounded above by $x_i$. Opening a station therefore only enlarges $\mathcal{Y}_t$. Every cut has slope $\beta = -\sum_k P_{jk}\pi_k \le 0$ (by the lower bound below), so $\hat{\mathcal{Q}}_{t+1}$ is also non-increasing in $x$, and the assumption holds by induction over the cuts generated.

Upper bound: For $\pi_i \ge \rho C_i$, the coefficient $\min(\rho C_i, \pi_i) = \rho C_i$ no longer depends on $\pi_i$, while $-\pi_i \hat{x}_i$ is non-increasing in $\pi_i$. Hence replacing $\pi_i$ by $\rho C_i$ never decreases $\mathcal{L}$.

Lower bound: Let $\pi_i < 0$, and let $\pi'$ equal $\pi$ except that $\pi'_i = 0$. If $\hat{x}_i = 0$: $\Lambda$ is non-decreasing in $\pi$, because the coefficients $\min(\rho C_i, \pi_i)$ are non-decreasing and $x \ge 0$, so $\mathcal{L}(\pi) \le \mathcal{L}(\pi')$. If $\hat{x}_i = 1$: under $\pi'$ the coefficient of $x_i$ is 0 and $h$ is non-increasing, so there is a minimiser $x^{**}$ of $\Lambda(\pi')$ with $x_i^{**} = 1$. Evaluating $\Lambda(\pi)$ at $x^{**}$ gives $\Lambda(\pi) \le \Lambda(\pi') + \pi_i$, so $\mathcal{L}(\pi) = \Lambda(\pi) - \pi^\top\hat{x} \le \Lambda(\pi') - \pi'^\top\hat{x} = \mathcal{L}(\pi')$. In both cases, negative multipliers do not improve the bound, which proves the proposition. Economically, $\pi_i$ is the value of entering the stage with station $i$ already open. It is non-negative because an open station can only reduce operating cost, and it cannot exceed the annualised cost $\rho C_i$ of opening the station in the current stage. The bounded box also keeps the cutting-plane model in the level method bounded.

### 3.3 Solution Algorithm

The solution algorithm is presented as pseudocode for the forward pass (Algorithm 1), the backward pass (Algorithm 2), the level method used to solve each Lagrangian dual (Algorithm 3), and the overall training and evaluation procedure (Algorithm 4). Setting each stage for the mid-path demand gives the deterministic rolling-horizon benchmark used in Section 5. In the pseudocode, $\mathcal{C}_{t,j}$ is the set of cuts for stage $t$ and demand state $j$, $\psi$ the epigraph variable of the expected cost-to-go, $\bar{\xi}$ an optional forced demand state, $\hat{x}$ the trial state and $\bar{\pi}_{t,k}$ the multiplier warm start. In Algorithm 3, $\boldsymbol{\mathcal{B}}$ is the bundle of evaluated points $u = (\pi_u, L_u, \sigma_u)$ with subgradients $\sigma_u$; $L^{\text{up}}$, $L_{best}$ and $L^{\text{lev}}$ are the cutting-plane upper bound, the best dual value found and the level; $\zeta \in (0,1)$ is the level weight and $m^{\max}$ the iteration limit. In Algorithm 4, $\nu$ indexes training iterations (at most $\nu^{\max}$) and $J$ is the number of evaluation replications.

*Algorithm 1: Pseudocode of the forward pass in the SDDiP solution procedure*

**Forward pass (one sampled demand path)**

```
Input: cuts {C_{t,j}}; transition matrix P; optional forced demand state ξ̄
x_0 ← 0;  ξ_1 ← Mid;  cost ← 0
for t = 1, …, T do
    if t > 1 then ξ_t ← ξ̄ if ξ̄ is given, else sample ξ_t ~ P(ξ_{t−1}, ·)
    solve min  c_t(x_{t−1}, x, y; ξ_t) + δψ
        s.t. x ≥ x_{t−1}, x ∈ {0,1}^n, y ∈ 𝒴_t(x, ξ_t), ψ ≥ 0, ψ ≥ γ_ℓ + β_ℓ^⊤ x  ∀ℓ ∈ C_{t+1,ξ_t} (ψ omitted if t = T)
      (time-limited MILP; the incumbent is a feasible decision)
    x_t ← x;  cost ← cost + δ^{t−1} c_t(x_{t−1}, x_t, y; ξ_t)
end for
return trial states (x_1, …, x_{T−1}), path (ξ_1, …, ξ_T), cost
```

*Algorithm 2: Pseudocode of the backward pass in the SDDiP solution procedure*

**Backward pass (Lagrangian cuts)**

```
Input: trial states (x_1, …, x_{T−1}) from Algorithm 1; multiplier warm starts π̄_{t,k}
for t = T, T−1, …, 2 do
   x̂ ← x_{t−1}
    for each demand state k ∈ {Low, Mid, High} do
        (Λ_k, π_k) ← LevelMethod(t, k, x̂, start = π̄_{t,k})
        π̄_{t,k} ← π_k
    end for
    for each parent demand state j ∈ {Low, Mid, High} do
        γ ← Σ_k P_{jk} Λ_k ;   β ← − Σ_k P_{jk} π_k
        add the cut ψ ≥ γ + β^⊤ x to C_{t,j}
   end for
end for
```

*Algorithm 3: Pseudocode of the level method in the SDDiP solution procedure*

**Algorithm 3: LevelMethod(t, k, $\hat{\mathbf{x}}$, $\boldsymbol{\pi}^0$):** maximise $\mathcal{L}_{t,k}(\pi;\hat{x})$ over $\pi \in [0,\rho C]$

evaluate $L^0 = \Lambda_{t,k}(\pi^0) - \pi^{0\top}\hat{x}$ (MILP; Λ from the solver's proven bound) and the subgradient $\sigma^0 = z^* - \hat{x}$,
  where $z_i^* = x_i^*$ if $\pi_i^0 < \rho C_i$, and $z_i^* = 0$ otherwise
  bundle $\mathcal{B} \leftarrow \{(\pi^0, L^0, \sigma^0)\}$; $(L_{best}, \pi_{best}) \leftarrow (L^0, \pi^0)$
  for $m = 1, \dots, m^{\max}$ do
    $L^{up} \leftarrow \max_{w,\,\pi\in[0,\rho C]} \{w: w \le L_u + \sigma_u^\top(\pi - \pi_u)\ \ \forall u \in \mathcal{B}\}$ (linear programming)
    if $L^{up} - L_{best} \le \varepsilon_{dual}$ then break
    level $L^{lev} \leftarrow L_{best} + \zeta(L^{up} - L_{best})$
    $\pi^m \leftarrow \arg\min_{\pi\in[0,\rho C]} \| \pi - \pi_{best} \|^2$ s.t. $L_u + \sigma_u^\top(\pi - \pi_u) \ge L^{lev}\ \ \forall u \in \mathcal{B}$ (quadratic programming)
    evaluate $L^m$ and $\sigma^m$ at $\pi^m$ as in line 1; $\mathcal{B} \leftarrow \mathcal{B} \cup \{(\pi^m, L^m, \sigma^m)\}$
    if $L^m > L_{best}$ then $(L_{best}, \pi_{best}) \leftarrow (L^m, \pi^m)$
  end for
return $\big(\Lambda_{t,k}(\pi_{best}), \pi_{best}\big) = (L_{best} + \pi_{best}^\top\hat{x},\ \pi_{best})$

*Algorithm 4: Pseudocode of training and policy evaluation in the SDDiP solution procedure*

**Algorithm 4: SDDiP training, lower bound and policy evaluation**

Initialise $\mathcal{C}_{t,j} \leftarrow \varnothing$ for all *t, j*; stall ← 0
for iteration $\nu = 1, \dots, \nu^{\max}$ do:
  run Algorithm 1 on a sampled path → trial states
  run Algorithm 2 on the trial states → new cuts
  $LB_\nu$ ← proven lower bound of the stage-1 problem with $\xi_1$ = Mid and cuts $\mathcal{C}_{2,\mathrm{Mid}}$
  if $|LB_\nu - LB_{\nu-1}|/|LB_{\nu-1}| \le \varepsilon_{LB}$ , then stall ← stall + 1 else *stall* ← 0
  if *stall* ≥ *patience* then break
end for
freeze the cuts; for replication $= 1, \dots, J$: run Algorithm 1 on a new random path → cost of replication
$UB \leftarrow \mathrm{mean}(cost) + 1.96 \times \mathrm{sd}(cost)/\sqrt{J}$;  $gap \leftarrow (UB - LB)/LB$
report the mid-, low- and high-branch policies: Algorithm 1 with ξ = Mid, Low, High

The cut computed for each child demand state $k$ in Algorithm 2 does not depend on the parent state. The same three dual solutions are therefore reused to update the cut sets of all three parent states, each with its own transition probabilities, at no additional computational cost. This is the standard treatment of Markovian uncertainty in SDDP-type algorithms (e.g. (Dowson and Kapelevich, 2021; Löhndorf and Shapiro, 2019; Philpott and de Matos, 2012)). With $T = 4$ and three demand states, the scenario tree has 40 nodes and 27 scenario paths and could still be solved as an extensive form. In this paper, SDDiP is used because its effort per iteration grows with the number of stages rather than with the number of scenario paths, which allows finer stage and branch resolution in larger applications.

As shown in Algorithm 4, the lower bound is the optimal value of the stage-1 problem with $\xi_1$= Mid and the current cuts, read from the solver's proven bound so that it remains valid under time limits. Training stops when the lower bound changes by less than $\varepsilon_{LB}$ (relative) in a given number of consecutive iterations (patience), or after $\nu^{\max}$ iterations. In this study, we assumed $\varepsilon_{LB} = 10^{-4}$ , patience = 3, $\nu^{\max} = 50$, $\varepsilon_{dual}$ = 100 AUD, $\zeta = 0.5$, $m^{\max} = 30$.

With the cuts then frozen, the policy is simulated on $J$ independently sampled demand paths (in this study, $J = 200$), using a random-number stream separate from training. The statistical upper bound is the sample mean plus the half-width of its 95% confidence interval, and the reported confidence bounds is $(\text{UB} - \text{LB})/\text{LB}$. Because time-limited forward solves return feasible incumbent decisions, the simulated policy is always feasible, and its expected cost is a valid upper bound on $Q_1$.

### 3.4 Energy Consumption Model

The network distance and drive time between each pair of nodes are calculated using the network analysis in ArcGIS Pro software. Energy consumption model: route distance and travel time between consecutive nodes are obtained from ArcGIS Pro network analysis. Each road centreline is discretised into contiguous segments with a nominal length of 1 km; the final segment retains its actual length. Elevation values were sampled along each route centreline from NASADEM, a NASA global digital elevation model at 1 arc-second (~30 m) horizontal resolution (NASA JPL, 2021).Elevation is sampled at the segment vertices. Segment grade is calculated as the endpoint elevation difference divided by planimetric length. Positive grade denotes uphill travel and negative grade denotes downhill travel in the direction assigned to the route. For each segment ($l$), the net resistive force acting on the truck ($F_l$) is computed from three components: the gravitational force along the grade, rolling resistance, and aerodynamic drag.

$$F_l = mass \times g \times \sin(\theta_l) + mass \times g \times C_r \times \cos(\theta_l) + 0.5\hbar_0 C_d A_f {v_l}^2 \tag{72}$$

where $mass$ is truck mass, $g$ gravitational acceleration, $C_r$ the rolling resistance coefficient, $C_d$ the aerodynamic drag coefficient, $A_f$ frontal area, $\hbar_0$ air density, and $v_l$ the travel speed on the segment (taken as the posted/design speed for that segment). The parameter $\theta_l = \arctan(grade_l)$ takes the road grade into account. Air density is treated as constant along the corridor rather than varying with sampled elevation, as a simplifying assumption appropriate to corridors without extreme elevation change. The sign of the net force determines whether the segment's traction draws energy from the battery or returns it. When the combined gravitational, rolling and aerodynamic resistance is positive, the motor must supply tractive effort and energy is drawn from the battery, subject to drivetrain efficiency losses. When a sufficiently steep descent makes the net force negative (the gravitational component exceeds rolling and aerodynamic resistance), the truck brakes regeneratively and returns energy to the battery, subject to the round-trip efficiency of the regenerative braking system.

In addition to traction energy, battery-electric trucks consume electrical energy to operate auxiliary systems such as cabin heating and cooling, battery thermal management, air compressors, power steering, lighting and onboard electronics. These loads are represented by a constant auxiliary power demand ($P_{aux}$), assumed to be independent of road grade and proportional to travel time over each segment. Consequently, the total battery energy required for a segment is the sum of the traction energy and the auxiliary energy consumption. This approach reflects the fact that auxiliary loads continue to draw power even on level roads or during regenerative braking events. The auxiliary energy consumption over the segment is calculated based on drive time on segment ($t_l$) and the assumed constant auxiliary electrical load ($P_{aux}$), as: $E_{aux,l} = P_{aux} t_l$. The traction energy ($E_{traction,l}$) for segment ($l$) is then calculated as:

$$E_{traction,l} = \begin{cases} \frac{F_l d_l}{h_{drive}}, & if\ F_l \geq 0 \\ F_l d_l h_{regenerative}, & if\ F_l < 0 \end{cases} \tag{73}$$

where $h_{drive}$ and $h_{regenerative}$ are the drivetrain and regenerative-braking efficiencies, respectively.

The total battery energy required for the segment ($e_l$) is therefore the sum of traction ($E_{traction,l}$) and auxiliary energy ($E_{aux,l}$). Because $E_{aux,l}$ is always positive, it partially offsets the energy recovered on regenerative segments, i.e. a descent yields net negative ($e_l$) (net energy returned to the battery) only once traction-side regeneration exceeds the auxiliary draw for that segment. Segment energy is then converted from joules to the same units as battery capacity (kWh) by dividing by 3,600,000.

Segment energies are aggregated over the complete route and over each leg between consecutive candidate facilities. The signed sum is used to identify the net energy requirement for each leg, subject to the model's treatment of regenerative braking and auxiliary load. This aggregation determines feasible facility spacing under the adopted battery and SoC assumptions. The parameters in Table 3 are held constant across routes and segments; payload, weather, traffic, acceleration and battery-temperature effects are not calibrated in the current implementation.

In this study, charging a battery from 0% to 100% SoC is assumed to take 90 minutes, while a swap takes 10 minutes. A battery removed during a swap is recharged to 100% at the station so that it can be used by other trucks. Each truck carries 4 packs of batteries, each charged at up to 120 kW in the 10–80% SoC range, the fastest part of the charging curve.

*Table 3: Parameter assumptions used in the segment-level energy model.*

| Symbol | Parameter | Assumed value | Description |
|---|---|---|---|
| $\boldsymbol{mass}$ | Loaded truck mass | 34,000 kg | Representative GVW for a heavy-duty battery-electric line-haul truck |
| $\boldsymbol{C_r}$ | Rolling resistance coefficient | 0.006 | Typical value for heavy truck tyres on paved highway |
| $\boldsymbol{C_d}$ | Aerodynamic drag coefficient | 0.60 | Typical value for a Class 8 tractor-trailer |
| $\boldsymbol{A_f}$ | Frontal area | 10.0 m2 | Representative frontal area for a loaded Class 8 tractor-trailer |
| $\hbar_{\mathbf{0}}$ | Air density | 1.225 kg/m3 | Standard atmosphere at 15C; treated as constant along the corridor (see Section 3) |
| $\boldsymbol{h_{drive}}$ | Drivetrain efficiency (motoring) | 0.90 | Combined motor, inverter and drivetrain efficiency while drawing power |
| $\boldsymbol{h_{regenerative}}$ | Regenerative braking efficiency | 0.60 | Round-trip efficiency of energy recovery during braking/descent |
| $\boldsymbol{P_{aux}}$ | Auxiliary electrical load | 4.0 kW | Representative constant auxiliary demand for a heavy-duty battery-electric truck (HVAC, battery thermal management, air compressor, steering, electronics) |

# 4 Data

The case study covers major freight corridors on Australia's east coast, connecting Brisbane, Sydney and Melbourne and including both travel directions. Candidate locations comprise (i) heavy-vehicle rest areas, (ii) existing passenger-EV charging sites and (iii) existing fuel-service stations. Only sites accessible in the direction of travel are retained, so the model does not require a U-turn or route detour. The candidate set is therefore a route-specific representation of feasible sites, not a complete national infrastructure inventory.

The demand for charging stations is calculated based on a percentage of the current flow on each corridor, considering future demand growth. The current flow, observed as average annual daily traffic (AADT) and the peak hour counts, is obtained from the counter stations alongside the highways, provided by the Commonwealth Government. Table 4 presents the route characteristics, peak-hour flow and the candidate stations alongside each route. It should be noted that the HV percentage might be significantly smaller than the values presented in Table 5; however, we assume the charging stations could be dual-purpose (for both passengers and trucks), and also the traffic volumes are more likely to grow significantly. Hence, we deliberately assume higher peak-hour volumes than the existing HV volumes on these corridors.

The fixed establishment cost represents the planning-level investment required to convert an existing site into a battery electric truck charging station. Notably, each charging facility needs to be planned and assessed as a bespoke item with detailed specifications on the battery energy storage system and grid supplies. In this study, we assume different candidate site categories would require different upgrade costs according to their anticipated level of existing infrastructure. The following assumptions are derived from the industry stakeholder consultation. Passenger EV charging stations are assumed to require the lowest incremental investment due to the presence of electrical infrastructure and utility approvals. The main upgrade costs for these candidate sites would be associated with a higher-capacity transformer, larger parking bays and a truck manoeuvring area. The existing gas service stations require the highest investment owing to the need for land acquisition, substantial electrical upgrades, a transformer, significant grid reinforcement and dedicated heavy-vehicle charging facilities. These sites are more likely to be private operators which already have commercial services, lighting, toilets, and sometimes a fast-food restaurant co-located, enabling long rest for truck drivers. Heavy vehicle rest areas require a moderate investment, since these sites already have truck parking, acceleration/deceleration lanes, and land is owned by the road infrastructure manager. These sites, however, would need new electrical connections, a transformer, a switchboard and charger foundations. For heavy vehicle rest areas, a separate Site Readiness Score is computed from existing amenities, assuming the weight parameters presented in Table 5, and discounting the fixed establishment costs.

*Table 4: Route characteristics and candidate stations, with assumed peak-hour demand*

| **Route Name** | **ID** | **Route** | **Direction *** | **Peak hourly demand** | **Total length (km)** | **Candidate charging/swapping stations** | | |
|---|---|---|---|---|---|---|---|---|
| | | | | | | **Passenger EV Charging stations** | **Gas stations** | **HV Rest areas** |
| New England highway | 1 | Newcastle to Brisbane | NB | 1065 | 793 | 1 | 49 | 11 |
| | 2 | Brisbane to Newcastle | SB | 654 | 589 | 1 | 40 | 8 |
| Pacific highway | 3 | Brisbane to Sydney | SB | 2230 | 1037 | 7 | 42 | 15 |
| | 4 | Sydney to Brisbane | NB | 2432 | 971 | 5 | 37 | 18 |
| Princes highway | 5 | Melbourne to Sydney | NB | 1641 | 1028 | 6 | 57 | 15 |
| | 6 | Sydney to Melbourne | SB | 1716 | 1046 | 4 | 52 | 15 |
| Hume highway | 7 | Melbourne to Sydney | NB | 693 | 870 | 2 | 14 | 39 |
| | 8 | Sydney to Melbourne | SB | 701 | 791 | 6 | 10 | 36 |
| Newell highway | 9 | Melbourne to Toowoomba | NB | 138 | 1533 | 3 | 48 | 15 |
| | 10 | Toowoomba to Melbourne | SB | 166 | 1577 | 3 | 38 | 9 |
| Great Western highway | 11 | Sydney to Dubbo | WB | 846 | 403 | 5 | 33 | - |
| | 12 | Dubbo to Sydney | EB | 452 | 403 | 3 | 24 | 3 |

** NB (Northbound); SB (Southbound); WB (Westbound); EB (Eastbound)*

*Table 5: Weight factors assumed for calculating the site readiness score for heavy vehicle rest areas.*

| **Facility** | Toilet | Lighting | Shelter | Water | Table | BBQ | Shade |
|---|---|---|---|---|---|---|---|
| **Weight** | 20 | 25 | 15 | 20 | 5 | 5 | 10 |

The annual fixed establishment costs are obtained by annualising the upfront cost with the capital recovery factor, using a discount rate of 6% and 20 years average lifetime. To focus on the trade-off between infrastructure deployment and freight service performance, both charging and battery swapping stations were assumed to have the same fixed establishment cost. This assumption reflects the strategic nature of the optimisation model, where the dominant investment components (land acquisition, site development, grid connection, transformers, and civil infrastructure) are assumed to be comparable across technologies. Technology-specific equipment costs, including charging units, battery inventories, and automated swapping systems, were excluded because they are highly uncertain, rapidly evolving, and vary considerably across manufacturers. Figure 1 presents all the potential sites considered as candidate charging stations. Notably, northbound/southbound and westbound/eastbound candidate sites are physically distinct locations.

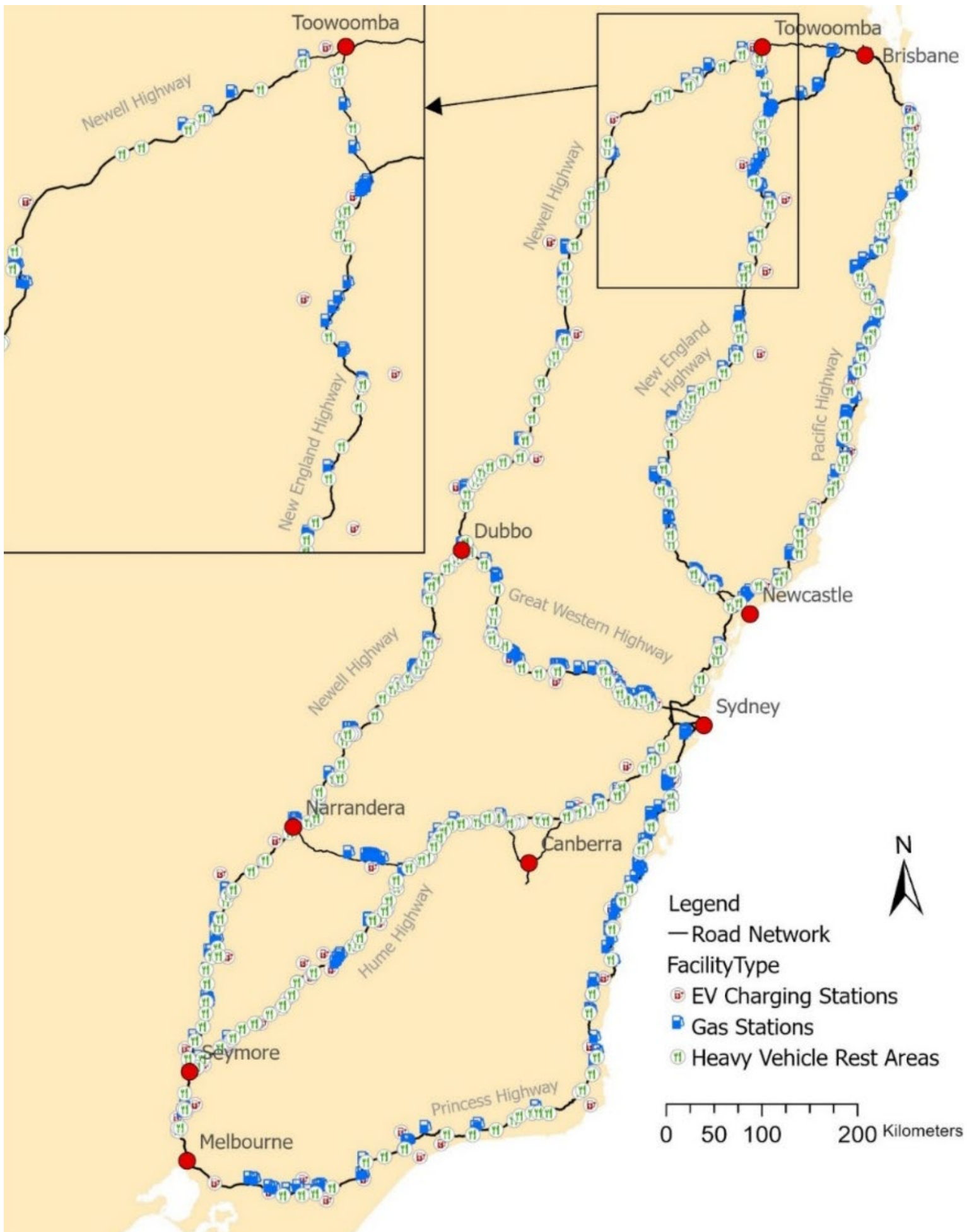

*Figure 1: Potential charging stations, including heavy vehicle rest areas, existing gas stations and passenger EV charging stations*

In Australia, heavy vehicle drivers must take mandatory rest under the Heavy Vehicle National Law (HVNL). Under its Standard Hours for solo drivers, a driver may work at most 5¼ hours in any 5½-hour period, 7½ hours in any 8 hours, 10 hours in any 11 hours and 12 hours in any 24 hours, with at least 15 minutes, 30 minutes, 60 minutes and 7 continuous hours of stationary rest in those periods, respectively (Heavy Vehicle (Fatigue Management) National Regulation, 2013). Stationary rest is time that a driver spends out of a heavy vehicle or in an approved sleeper berth of a stationary fatigue-regulated heavy vehicle, so the mandatory rest can be used for charging or battery swapping (opportunity charging). This study uses a stylised two-tier rule that requires more frequent rest than these minimums: a 10-minute short break after at most 2 hours of driving and a 30-minute rest after at most 5 hours of driving, following the NSW Heavy Vehicle Driver Handbook (Government of New South Wales). The 24-hour limit and its 7-hour rest are not modelled (Section 7). We assume that charging is unattended (the driver plugs in and then rests away from the vehicle) and that swapping is automated; the few minutes needed to connect or position the vehicle are neglected, and the remaining dwell counts as rest. Because refuelling counts as work under the HVNL, this assumption should be confirmed for attended charging.

# 5 Experimental Results

This section reports the computational performance and station-siting results of the model variants introduced in Section 3, solved as MILP problems across all 12 routes. The model formulation assumes that, at each en-route station, a trucking operator chooses either to charge all onboard battery packs simultaneously or to swap one or more of them. Although the proposed formulation in Section 3 handles both charging and swapping decisions within a single integrated model, the two energy-replenishment strategies are treated separately in the experimental analysis to more closely reflect current operational practice. This separation is motivated by existing battery-swapping schemes deployed globally, in which operators participating in swapping-based business models rely predominantly on battery exchange rather than independently charging onboard batteries. Such schemes also mitigate operator concerns related to battery ownership, degradation, and residual value, as battery management and replacement are typically handled within the swapping ecosystem. Commercial precedents for both modular multi-pack swapping and the exclusive adoption of a single energy-replenishment strategy are already well established. Battery-as-a-service subscriptions, first offered for passenger cars, have since been extended to freight. CATL's Qiji Energy division has deployed a modular heavy-truck swapping system in which one to three standardised 171-kWh battery modules can be assembled per truck, with users selecting the number of batteries according to their needs and optimising swap routes through a dedicated cloud reservation platform (Barnard, 2026). Automated multi-pack exchange has also been demonstrated for heavy-duty trucks outside China. In Germany's eHaul project, a swap robot approaches the truck from the left and right, lifting and removing the battery packs before retrieving and reinserting fully charged replacements, with the entire process completed in approximately ten minutes (Technische Universität Berlin, 2026).

These practices are also reflected in the academic literature. Wang et al. (2025) directly compare battery-swapping and fast-charging infrastructure for electric heavy-duty trucks, treating them as two distinct network designs rather than a hybrid system. Their evaluation finds that battery-swapping stations significantly increase transportation efficiency by reducing vehicle downtime relative to fast charging, although they may require larger battery inventories, and that swapping stations using medium-sized batteries offer the best balance of time and labour efficiency. These practical applications and studies collectively motivate us to assume that a trucking operator adopts either a charging strategy or a swapping strategy exclusively, rather than combining both. These service models are treated as mutually exclusive from the perspective of individual trucks: trucks subscribed to the charging network use only public charging stations, while trucks subscribed to the battery-swapping network use only swapping stations. Therefore, the two infrastructure networks are modelled separately, while their spatial deployment may be jointly optimised. However, we provide the integrated model framework capable of handling both swapping vs. charging decisions simultaneously, for readers who wish to study a combined business model.

To model the demand uncertainty, four demand stages (5%, 10%, 15%, 20% of BET penetration), with three stochastic demand branches (low, mid, and high) are assumed. The demand tree is modelled as a Markov chain, in which each subsequent demand stage branches three ways from its parent. The first node represents the Stage 1 demand (5% BET penetration), and then propagates into 40 distinct nodes (Stage 1: 1; Stage 2: 3 demand scenarios; Stage 3: 9 demand scenarios; and Stage 4: 27 demand scenarios). Table 6 presents the stochastic demand transition matrix (i.e. branching probabilities). In the absence of real BET adoption data, these demand transition values are hypothesised for methodological demonstrations. The rationale behind the assumed values is as follows. The diagonal dominance (i.e. higher values assumed for diagonal cells) enforces that staying in the same state is always most likely (i.e. model adoption momentum). In other words, technology adoption processes tend to persist. For example, if demand has been growing slowly in this stage, it is more likely to keep growing slowly next stage than to suddenly swing to a completely different regime. Looking at the off-diagonal pattern within each row, the values are assumed to follow an ordered scale, encoding the assumption that demand shifts tend to happen gradually. For example, a low adoption trajectory drifting up to moderate adoption is far more plausible in one stage than leaping straight to high adoption. This assumption attempts to mirror diffusion-of-technology curves (S-curve, not step function). The low and high scenarios are assumed symmetric, indicating the absence of directional bias, meaning that the model has no prior view on whether demand is more likely to accelerate than decelerate. The mid row is a symmetric 25/50/25 split, so from the mid branch the next-stage outcome is equally likely to move up or down. The branch multipliers are 0.7, 1.0 and 1.3 times the mid-path demand.

*Table 6: Demand transition matrix assumed in this study.*

| **From/To** | **Low** | **Mid** | **High** |
|---|---|---|---|
| **Low** | 0.5 | 0.40 | 0.10 |
| **Mid** | 0.25 | 0.50 | 0.25 |
| **High** | 0.10 | 0.40 | 0.50 |

Table 7 presents the number of decision variables for the swap-only and charge-only models for each route for the whole demand tree.

*Table 7: Number of decision variables in the SDDiP model*

| **Route** | **Total Variables** | **Binary Variables** | **Integer Variables** | **Continuous Variables** |
|---|---|---|---|---|
| **1** | 85,360 | 32,600 | 4,880 | 47,880 |
| **2** | 67,680 | 25,840 | 3,840 | 38,000 |
| **3** | 89,440 | 34,160 | 5,120 | 50,160 |
| **4** | 84,000 | 32,080 | 4,800 | 47,120 |
| **5** | 108,480 | 41,440 | 6,240 | 60,800 |
| **6** | 98,960 | 37,800 | 5,680 | 55,480 |
| **7** | 79,600 | 30,360 | 4,400 | 44,840 |
| **8** | 73,120 | 27,920 | 4,160 | 41,040 |
| **9** | 95,760 | 36,520 | 5,280 | 53,960 |
| **10** | 75,200 | 28,640 | 4,000 | 42,560 |
| **11** | 54,080 | 20,640 | 3,040 | 30,400 |
| **12** | 43,200 | 16,480 | 2,400 | 24,320 |

All optimisation models were implemented in Python using the Gurobi Optimizer (version 13.0.2) via its gurobipy interface (Gurobi Optimization, 2024). Models were solved on high computational infrastructure, using 16 CPU cores and 256 GB RAM per job. Each route's SDDiP solve proceeds in two phases: a training phase (alternating forward and backward passes, building Benders-style cuts via a level-method Lagrangian dual) that stops once the lower bound stalls (relative change <= 1e-4 for 3 consecutive iterations), followed by an evaluation phase that draws J independent forward-pass replications (J = 200 by default) against the fixed, trained cuts to produce a statistical upper bound (mean + 95% confidence-interval half-width), a materially more reliable convergence signal than comparing the lower bound against any single sampled path (see Section 3 and Algorithm 4). A route was classed as converged once the gap between this statistical upper bound and the proven lower bound fell to 5% or below. High demand long routes (e.g. routes 3, 4 and 5) initially closed this gap slowly under the baseline solver configuration. Diagnostic inspection of the candidate-station data revealed that a large share of candidate sites on these corridors share identical fixed installation costs and identical composite site-suitability scores (e.g. 42 of 64 candidates on Route 3 and 37 of 60 on Route 4 are tied on both cost and score). This is a well-documented pathology in facility-location-type mixed-integer programs, where many candidate locations are cost- and score-equivalent, the solver cannot distinguish between combinatorially many interchangeable solutions, which weakens the LP relaxation bound and slows branch-and-bound convergence independently of the underlying decision problem's genuine difficulty. To mitigate this, Gurobi's Symmetry parameter was set to its most aggressive detection-and-breaking level (Symmetry = 2) for these routes, directing the solver to spend additional presolve effort identifying and breaking these symmetric candidate-site structures before branching begins. With this setting, all swap-only routes except Route 9 (5.35%) and seven of the twelve charge-only routes reached the 5% target; the remaining policy-evaluation confidence bounds are reported in Table 10.

## 5.1 Station-siting Decisions in SDDiP against Deterministic Rolling Horizon Model

Table 8 and Table 9 report the station-siting decisions for the battery-swap and charge network, respectively. Some high-traffic routes required a higher number of battery inventories (in the swap-only model) or chargers (in the charge-only model).

*Table 8: Results for the SDDiP swap-only model by routes and demand stages.*

| **Swap-only model** | | | | | |
|---|---|---|---|---|---|
| **Route Name** | Direction | Route ID | Stations built in each stage | Cap battery inventory at each station | Maximum utilisation rate |
| **New England highway** | NB | 1 | 3→3→3→3 | 200 | 18%→36%→53%→71% |
| | SB | 2 | 3→3→3→3 | 200 | 11%→22%→33%→44% |
| **Pacific highway** | SB | 3 | 3→3→7→7 | 200 | 37%→74%→98%→93% |
| | NB | 4 | 3→3→3→6 | 200→200→250→300 | 41%→81%→97%→95% |
| **Princes highway** | NB | 5 | 4→4→4→4 | 200→200→250→300 | 27%→55%→66%→73% |
| | SB | 6 | 4→4→4→8 | 200 | 29%→57%→86%→86% |
| **Hume highway** | NB | 7 | 4→4→4→4 | 200 | 12%→23%→35%→46% |
| | SB | 8 | 3→3→3→3 | 200 | 12%→23%→35%→47% |
| **Newell highway** | NB | 9 | 7→7→7→7 | 200 | 2%→5%→7%→9% |
| | SB | 10 | 6→6→6→6 | 200 | 3%→6%→8%→11% |
| **Great Western highway** | WB | 11 | 1→1→1→1 | 200 | 14%→28%→42%→56% |
| | EB | 12 | 1→1→1→1 | 200 | 8%→15%→23%→30% |

The maximum number of chargers per station was set by route and stage (Table 9), from 50 on the Newell Highway and the eastbound Great Western Highway to 300 on the Pacific Highway, and was increased where a lower cap made a route infeasible. The resulting peak utilisation ranges from 28% to 100%. Where the cap is held fixed while demand grows, as on the Newell Highway, utilisation rises from 28–34% in the second stage to 56–66% in the final stage.

*Table 9: Results for SDDiP charge-only model by routes and demand stages.*

| Charge-only model | | | | | |
|---|---|---|---|---|---|
| **Route Name** | Direction | Route ID | Stations built in each stage | Assumed chargers required at each station | Maximum utilisation rate |
| **New England highway** | NB | 1 | 8→8→8→8 | 50→100→150→200 | 54%→53%→53%→53% |
| | SB | 2 | 4→4→4→4 | 50→100→150→200 | 66%→65%→65%→66% |
| **Pacific highway** | SB | 3 | 13→13→13→13 | 50→100→200→300 | 74%→74%→84%→74% |
| | NB | 4 | 12→12→12→12 | 50→100→200→300 | 82%→81%→91%→81% |
| **Princes highway** | NB | 5 | 9→10→10→10 | 50→100→150→200 | 82%→82%→82%→82% |
| | SB | 6 | 10→10→10→10 | 50→100→150→200 | 86%→86%→86%→86% |
| **Hume highway** | NB | 7 | 7→7→7→7 | 50→100→150→200 | 100%→69%→69%→70% |
| | SB | 8 | 4→7→7→7 | 50→70→100→150 | 70%→50%→53%→93% |
| **Newell highway** | NB | 9 | 9→9→9→9 | 50→50→50→50 | 100%→28%→42%→56% |
| | SB | 10 | 8→8→8→8 | 50→50→50→50 | 89%→34%→50%→66% |
| **Great Western highway** | WB | 11 | 2→3→3→3 | 50→70→100→150 | 84%→60%→63%→57% |
| | EB | 12 | 2→2→3→3 | 50→50→50→50 | 46%→90%→68%→90% |

Figure 2 presents the stations built and the maximum number of required battery inventories in every demand stage at the mid-path demand branch by route at every demand stage, showing how the number of stations built varies across the network, from the lowest-flow corridors (e.g. route IDs 1-2 & 7-12) to the busiest routes (e.g. route IDs 3-6). Routes 9 and 10 need lower battery inventories but a relatively higher number of stations to satisfy the peak-hour demand compared to other routes. This is due to the geographic stretch of the candidate stations along the routes and the fact that seven and six swapping stations will be needed for routes 9 and 10, respectively, to satisfy the charging needs, regardless of the lower route demands.

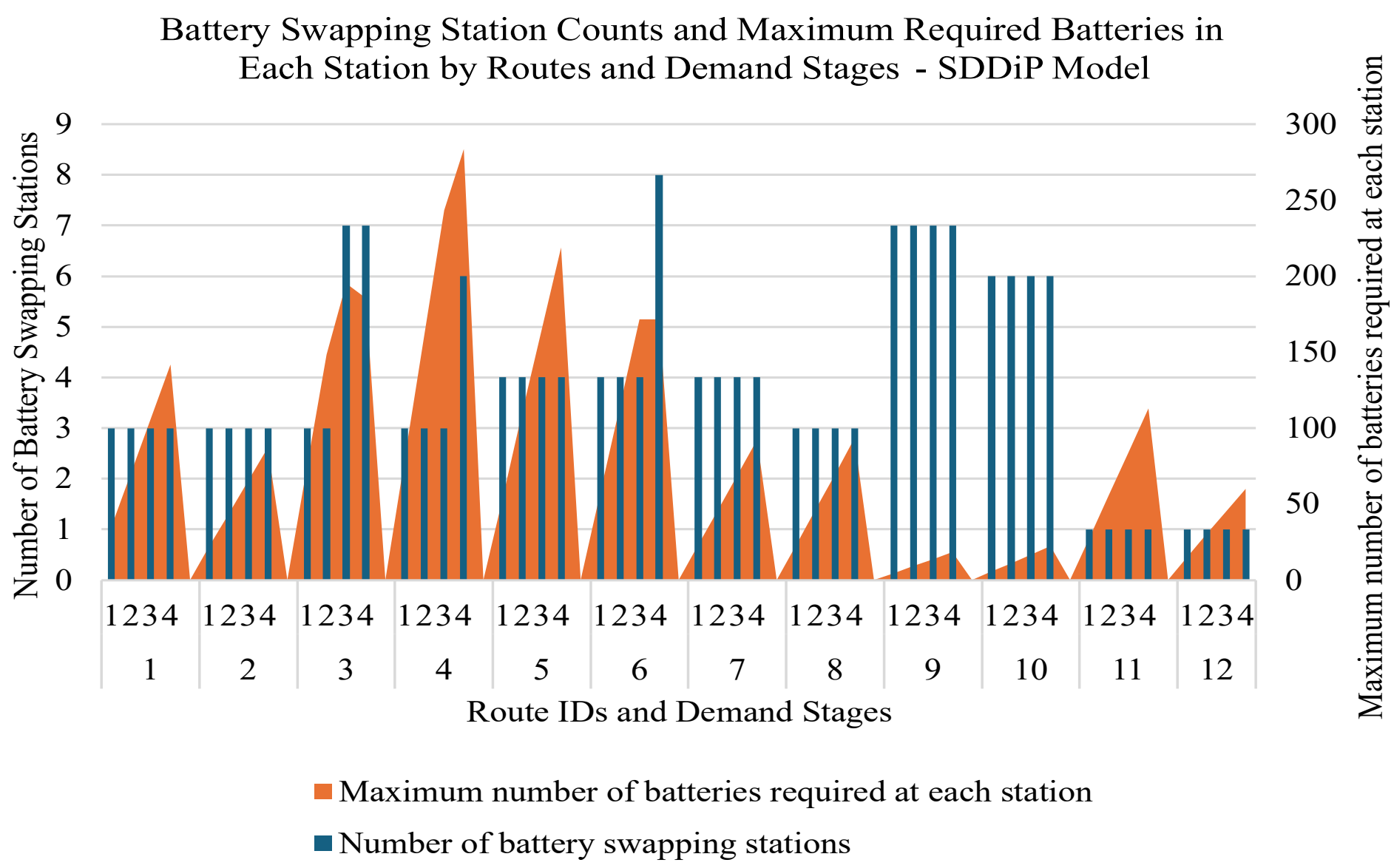

*Figure 2: Results of battery swapping station counts and maximum required number of batteries in the SDDiP model by routes and demand stages.*

Figure 3(a-c) shows the number of charging stations and the peak number of chargers required at the busiest station on each route along the low, mid and high demand branches of the charge-only SDDiP model. Each branch traces the policy's decisions when demand remains on that branch in every stage, at 70%, 100% and 130% of the mid-path penetration, respectively.

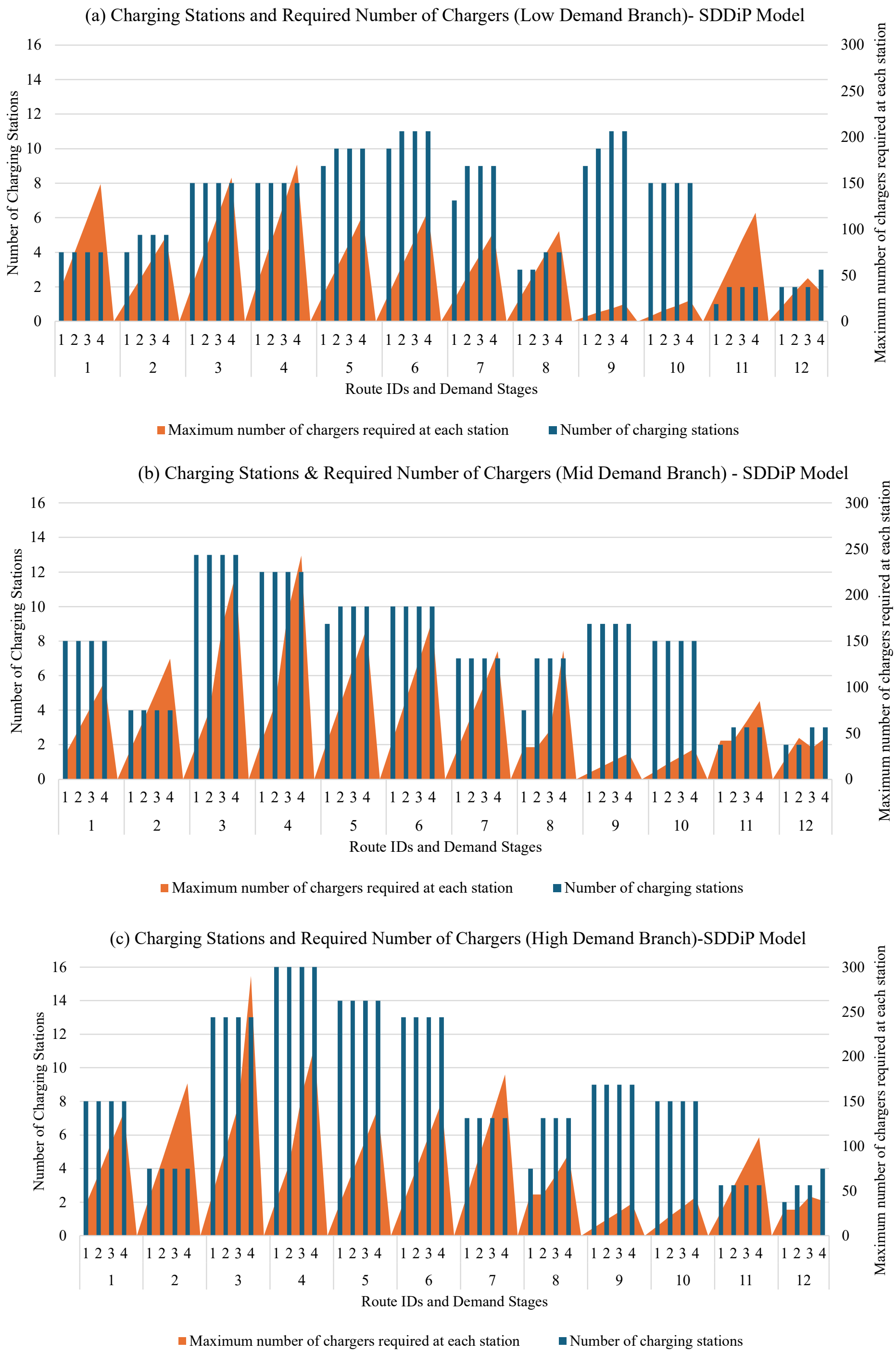

*Figure 3: Results of charging station counts and maximum required number of chargers at each station by route and demand stage.*

The size of the network scales with demand, but less than proportionally. By the final stage, the model builds 83, 94 and 106 stations on the low, mid and high branches, at capital costs of AUD 1,148M, 1,262M and 1,475M. The ±30% variation in demand therefore produces only a −12% to +13% change in station count and a −9% to +17% change in capital cost. The remaining demand is absorbed by installing more chargers at existing sites and using them more intensively, where the average peak utilisation of the busiest station on each route rises from 57% on the low branch to 73% on the mid branch and 77% on the high branch. On every branch, the build-out is concentrated in the early stages, with more than 95% of the final-stage stations already in place by the second stage, so hedging against uncertainty acts mainly on which sites are selected rather than on when they are built. The peak charger requirement, however, does not increase monotonically with demand. Across the network, the busiest station needs 170, 243 and 290 chargers in the final stage, but in the third stage the mid branch (182) exceeds the high branch (158). On Routes 3, 4 and 8, the high branch builds additional stations that spread the demand, lowering the load at the busiest site; on Route 4 in the final stage, for example, 16 stations with a peak of 211 chargers replace 12 stations with a peak of 243. On the low-volume Routes 9 and 10 the number of stations barely changes across branches (8–11), indicating that siting there is driven by the driving range between stations rather than by charging capacity, and only the number of chargers scales with demand.

The SDDiP model was benchmarked against a myopic stage-by-stage solve with irreversible stations, chargers and batteries. The same 12-route network was also solved with a deterministic rolling-horizon model, which re-solves each stage as a single mixed-integer program against one point-forecast demand value, with no representation of demand uncertainty. Figure 4(a-b) present the maximum required stations, battery inventory and chargers at each station in both swap-only and charge-only model by route and demand stage. Figure 4(a-b) can be compared to Figure 2 and Figure 3(b) under the SDDiP model along its mid-demand path, which follows the same demand trajectory as the deterministic case. For the charge-only network, the two models are closely aligned. They build almost the same number of stations in every stage (87 vs 88 at 5% penetration and 95 vs 94 at 20%), and the peak charger requirement is identical on every route and stage, reaching 243 chargers on Route 4 at 20% penetration. The models differ mainly in which sites they select. Only 67 of the stage-4 SDDiP stations (71%) coincide with the deterministic solution, with the largest differences on Routes 1, 2, 7, 8 and 12. The SDDiP sites are also more expensive, raising the capital cost by 8.2% in the first stage and 4.4% in the final stage.

The swap-only network diverges more strongly. SDDiP builds fewer stations once demand grows (46 vs 52 at 15% and 53 vs 59 at 20% penetration), and only about half of its stage-4 stations coincide with the deterministic ones, with entirely different sites on Routes 3 and 4 in the first two stages. Concentrating demand on fewer sites raises the peak battery requirement at the busiest station, from 162 to 284 batteries on Route 4 and from 191 to 219 on Route 5 at 20% penetration. The cost profile is also shifted forward. The SDDiP model spends 5.8% more than the deterministic model in the first two stages but 3.0% less by the final stage. These differences are consistent with the SDDiP policy anticipating the high-demand branch. The rolling-horizon model optimises each stage only for the demand observed at that stage and adds stations incrementally as demand grows; Route 3 swap stations, for instance, rise from 5 to 10. The SDDiP model instead commits early to sites and capacities that remain cost-effective across the possible future demand paths, even at the cost of higher early investment, and so ends with a smaller and more concentrated network.

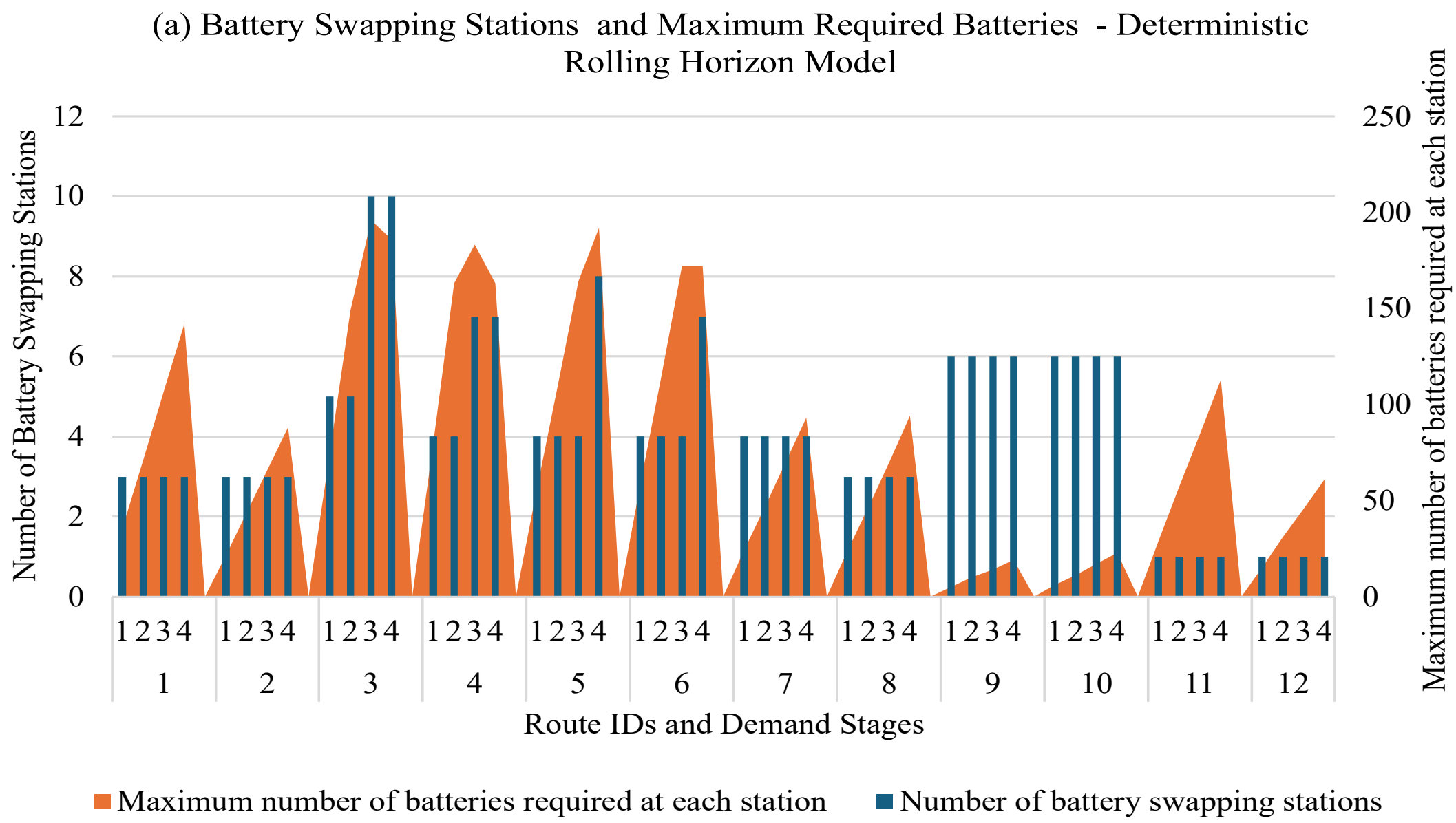

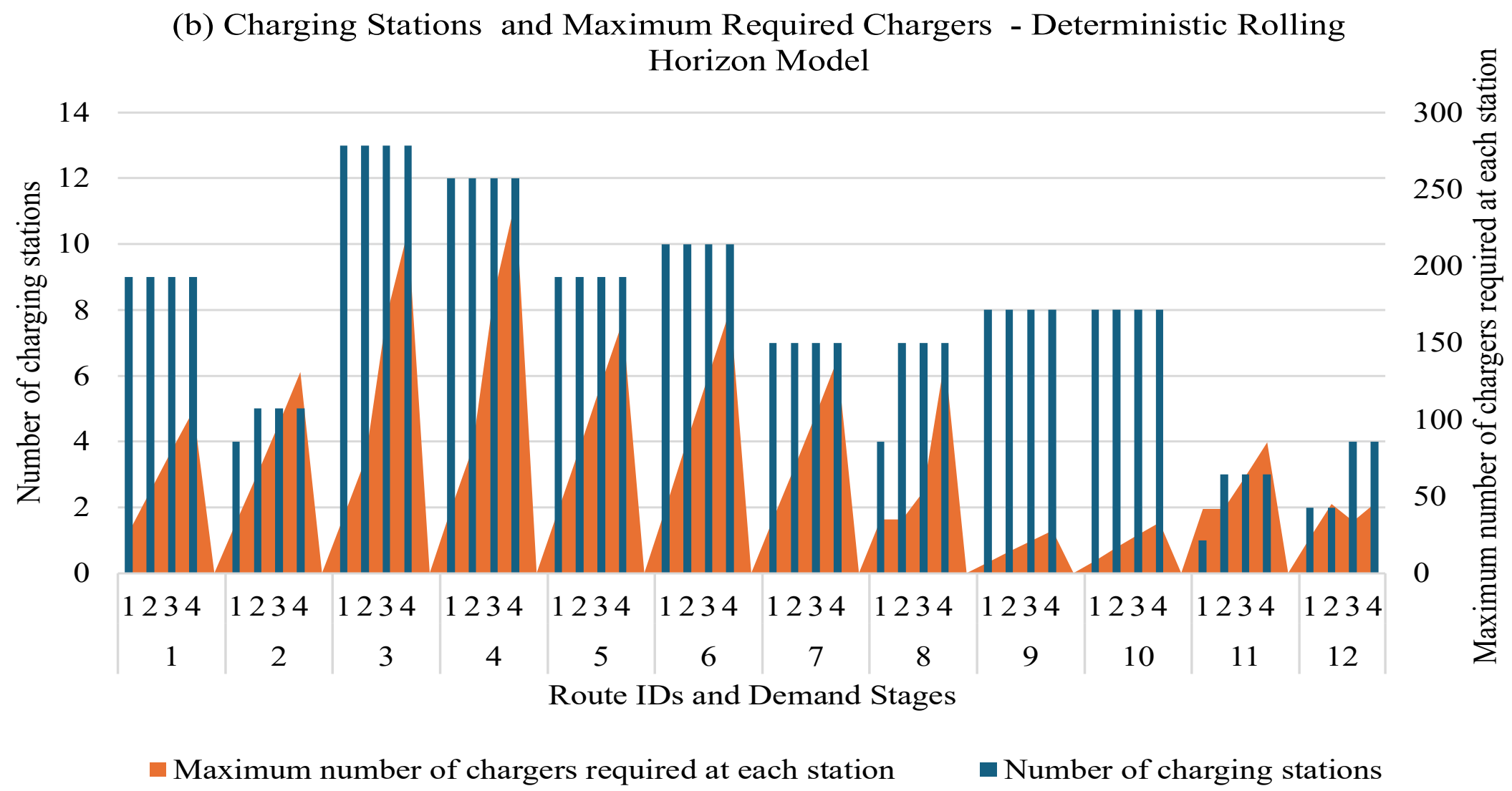

*Figure 4: Maximum station counts and required number of batteries and chargers in the deterministic rolling horizon model, by routes and demand stages.*

Table 10 reports the lower bound (LB) and statistical upper bound (UB) of the SDDiP objective for each route under the swap-only and charge-only models. The SDDiP objective value represents the total expected discounted cost of one multi-stage stochastic decision problem, which includes annualised CAPEX and VoT-monetised delay, summed across all 4 demand stages with per-stage discounting, under genuine demand uncertainty. The LB is a deterministic, valid bound on the minimum expected discounted cost, obtained by solving the first-stage problem with the cost-to-go function approximated by the Lagrangian cuts accumulated during training. The UB is a statistical estimate of the expected cost of the trained policy, where the policy is fixed and simulated over independently sampled demand paths, and the UB is the sample mean plus the half-width of its 95% confidence interval. The policy-evaluation confidence bound, (UB − LB)/LB, therefore bounds how far the trained policy may be from the true optimum, but it also includes the sampling error of the UB estimate. The reported relative difference therefore combines policy suboptimality with sampling uncertainty and should not be read as a deterministic proof gap.

*Table 10: Lower bound (LB) and upper bound (UB) objective function values for swap-only and charge-only models*

| Route Name | Direction | Route ID | Swap-only model | | | Charge-only model | | |
|---|---|---|---|---|---|---|---|---|
| | | | Objective Function Value (million AUD) | | Confidence bound (%) | Objective Function Value (million AUD) | | Confidence bound (%) |
| | | | LB | UB | | LB | UB | |
| New England highway | NB | 1 | 100.62 | 103.55 | 2.91 | 359.41 | 387.95 | 7.94 |
| | SB | 2 | 88.86 | 91.34 | 2.8 | 205.30 | 211.61 | 3.07 |
| Pacific highway | SB | 3 | 400.22 | 413.84 | 3.4 | 1127.05 | 1157.87 | 2.73 |
| | NB | 4 | 446.96 | 463.48 | 3.69 | 1460.70 | 1513.27 | 3.60 |
| Princes highway | NB | 5 | 265.18 | 273.04 | 2.96 | 899.89 | 933.01 | 3.68 |
| | SB | 6 | 282.28 | 293.69 | 4.04 | 898.78 | 931.91 | 3.69 |
| Hume highway | NB | 7 | 266.15 | 274.42 | 3.11 | 330.73 | 357.12 | 7.98 |
| | SB | 8 | 101.55 | 105.32 | 3.71 | 249.83 | 259.92 | 4.04 |
| Newell highway | NB | 9 | 67.18 | 70.77 | 5.35 | 101.85 | 106.47 | 4.54 |
| | SB | 10 | 67.58 | 69.35 | 2.62 | 101.39 | 107.59 | 6.12 |
| Great Western highway | WB | 11 | 37.41 | 38.53 | 2.98 | 87.99 | 92.84 | 5.52 |
| | EB | 12 | 16.65 | 17.09 | 2.63 | 42.53 | 44.20 | 3.79 |

*Note: The lower bound (LB) is a certified bound from the trained value function; the statistical upper bound (UB) is a confidence-interval estimate from policy evaluation. Together they bracket the optimal expected cost, up to the sampling error of the UB.*

The swap-only model converges tightly and consistently: policy-evaluation confidence bounds range from 2.62% to 5.35% (mean 3.35%), and only route 9 exceeds 5%. The charge-only model yields confidence bounds of 3.07–7.9%, with four routes above 5%. The largest are on the New England Highway northbound (7.94%) and the Hume Highway northbound (7.98%). The looser convergence of the charge-only model is consistent with its harder stage subproblems. Charger occupancy depends on continuous charging durations coupled to battery state-of-charge dynamics, and demand is split across sub-flows of drivers, so the mixed-integer stage problems are larger and more often terminated at their time limit. In such cases, the cuts are built from the solver's proven bound rather than the optimal value, which weakens the cost-to-go approximation and keeps the LB lower. In addition, the charge-only objective values are 1.2–3.6 times larger than their swap-only counterparts because charging adds substantially to driver delay. Delay is more sensitive to the demand realisation than station capital cost, which widens the spread of simulated policy costs and hence the confidence interval on the UB. Even so, ten of the twelve charge-only routes converge to within about 6%, indicating that the SDDiP policies are near-optimal for both technologies, with the charge-only results on routes 1 and 7 warranting more cautious interpretation.

Table 11 reports each route's computational time comparison between the deterministic aand the SDDiP models. For the deterministic rolling-horizon model, the computation time is reported for the sum of the four per-stage MILP solve times (each solved once, to a proven exact optimum). For SDDiP, the computational time is the reported end-to-end elapsed time covering the full SDDiP training loop (repeated forward and backward passes until the lower bound stalled or the iteration cap was reached) plus the J-replication policy evaluation used to compute the statistical upper bound. Every deterministic rolling horizon stage/route solve reached a proven global optimum (Gurobi status 2, MIP gap 0%) in well under ten minutes, reflecting the much smaller size of a single-scenario problem compared to SDDiP's multi-branch scenario tree. SDDiP, by contrast, solves a low/mid/high demand scenario tree at every stage and hedges the station-siting decision against the probability-weighted range of future demand outcomes, at the cost of far greater computational effort. The computational time difference is even more pronounced in the charge-only model for high-traffic routes, such as the Pacific and Princes Highways.

*Table 11: Computational time (min) per iteration for the SDDiP model vs. the deterministic rolling-horizon model.*

| ***Route Name*** | ***Direction*** | **Route** | **Swap-only model** | | | **Charge-only model** | | |
|---|---|---|---|---|---|---|---|---|
| | | | Rolling-horizon | SDDiP | SDDiP / Rolling-horizon | Rolling-horizon | SDDiP | SDDiP / Rolling-horizon |
| New England highway | NB | 1 | 3 | 218 | 64× | 4 | 680 | 164× |
| | SB | 2 | 1 | 66 | 65× | 1 | 12 | 33× |
| Pacific highway | SB | 3 | 51 | 3090 | 60× | 3620 | 39738 | 10× |
| | NB | 4 | 34 | 5262 | 154× | 3702 | 51048 | 13× |
| Princes highway | NB | 5 | 111 | 782 | 6× | 4 | 19320 | 5336× |
| | SB | 6 | 35 | 3484 | 99× | 20 | 22249 | 1134× |
| Hume highway | NB | 7 | 5 | 427 | 90× | 2 | 143 | 70× |
| | SB | 8 | 2 | 370 | 148× | 30 | 26747 | 891× |
| Newell highway | NB | 9 | 10 | 569 | 58× | 1 | 80 | 90× |
| | SB | 10 | 5 | 392 | 85× | 1 | 22 | 47× |
| Great Western highway | WB | 11 | 1 | 133 | 266× | 1 | 8 | 18× |
| | EB | 12 | 1 | 27 | 80× | 1 | 1261 | 1400× |

## 5.2 Driver Hours of Service and Charging Schedule

The model determines the driving schedule for every route, and the charging (or swapping) schedule for every sub-flow, in each demand stage and for both the charge-only and swap-only models. Figure 5 shows the driver schedule on Route 1 (Newcastle–Brisbane via the New England Highway, northbound) in the first demand stage (5% BET penetration) under the SDDiP solution. The figure reports the cumulative arrival and departure times from the origin, the dwell time and the battery SoC at each stop, and whether the stop is a short break, a long rest, or a break combined with a charging event. At 5% penetration, the route's peak-hour BET flow exceeds the 50-charger capacity of a single station, so the model splits the flow into two subgroups of drivers that recharge at different stations (using equation 31). Both subgroups follow the same timetable, completing the 718-minute trip with six stops: three short breaks and five long rests, with a combined dwell of 244 minutes. They alternate charging sites, however. Every charging event coincides with a mandatory short break or long rest, so charging is absorbed into rest time required under the hours-of-service rules rather than adding separate stops. Where the charge takes longer than the required break, the dwell is extended. Batteries are typically recharged from 10–38% SoC to the 80% charging cap, taking 11–42 minutes per event. It should be noted that a long rest does not necessarily coincide with charging to the maximum SoC. A battery is charged only as much as is needed to reach the next charging stop, or the destination, while keeping the minimum SoC reserve; on Route 1, for example, the batteries charged during long rests reach 74–78% rather than the 80% cap at two stops, and some long rests involve no charging at all. Both subgroups reach Brisbane at the 10% minimum SoC, showing that the schedule uses the available battery range fully without violating the reserve requirement.

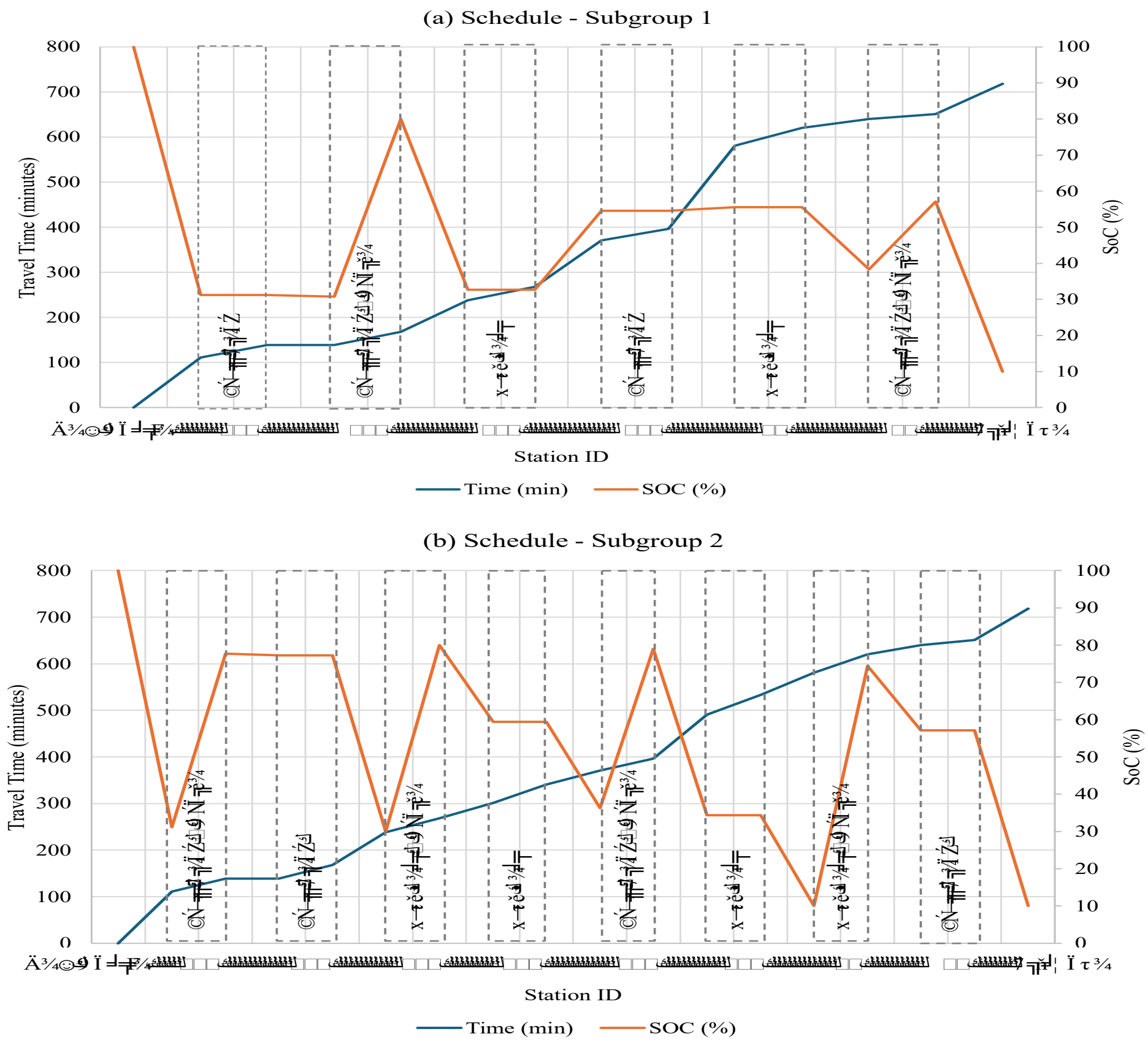

*Figure 5: Illustrative example of rest and charging schedule on route 1.*

## 5.3 Cost–delay Trade-off: Single-objective Extremes

To test how much of each route's station-siting decision is driven by capital cost rather than driver delay, the deterministic rolling-horizon model was re-solved under two single-objective extremes with all capacity and HoS constraints unchanged: minimising station CAPEX alone (cost-only, VoT = 0) and minimising driver delay alone (delay-only). Figure 6 compares the stations built in the final stage (20% penetration) under the two extremes for the swap-only (a) and charge-only (b) models. The cost-only results show the least-cost network that satisfies the range, capacity and HoS constraints.

Across the 12 routes, the delay-only objective required more charging stations than the cost-only objective (range: 2×–7×), resulting in a significant increase in total station CAPEX when built to the delay-minimising configuration. For swapping the increase is larger still (6–38×): with no cost penalty, every additional station shortens some dwell, so the delay-only model opens almost every candidate site. Comparing the combined-objective results of the deterministic model (Figure 4) with the single-objective results in Figure 6 shows that, in the final stage, the cost-only network coincides with the combined-objective network on every charge-only route and on 11 of the 12 swap-only routes. At the assumed value of time of 3 AUD per truck-minute, the least-cost network is therefore already the cost–delay optimum, and the cost–delay trade-off is inactive. This is consistent with charging being absorbed into mandatory rest (Section 5.2), which leaves little delay for additional stations to remove; a value-of-time sweep would show at what value of time additional stations begin to pay off.

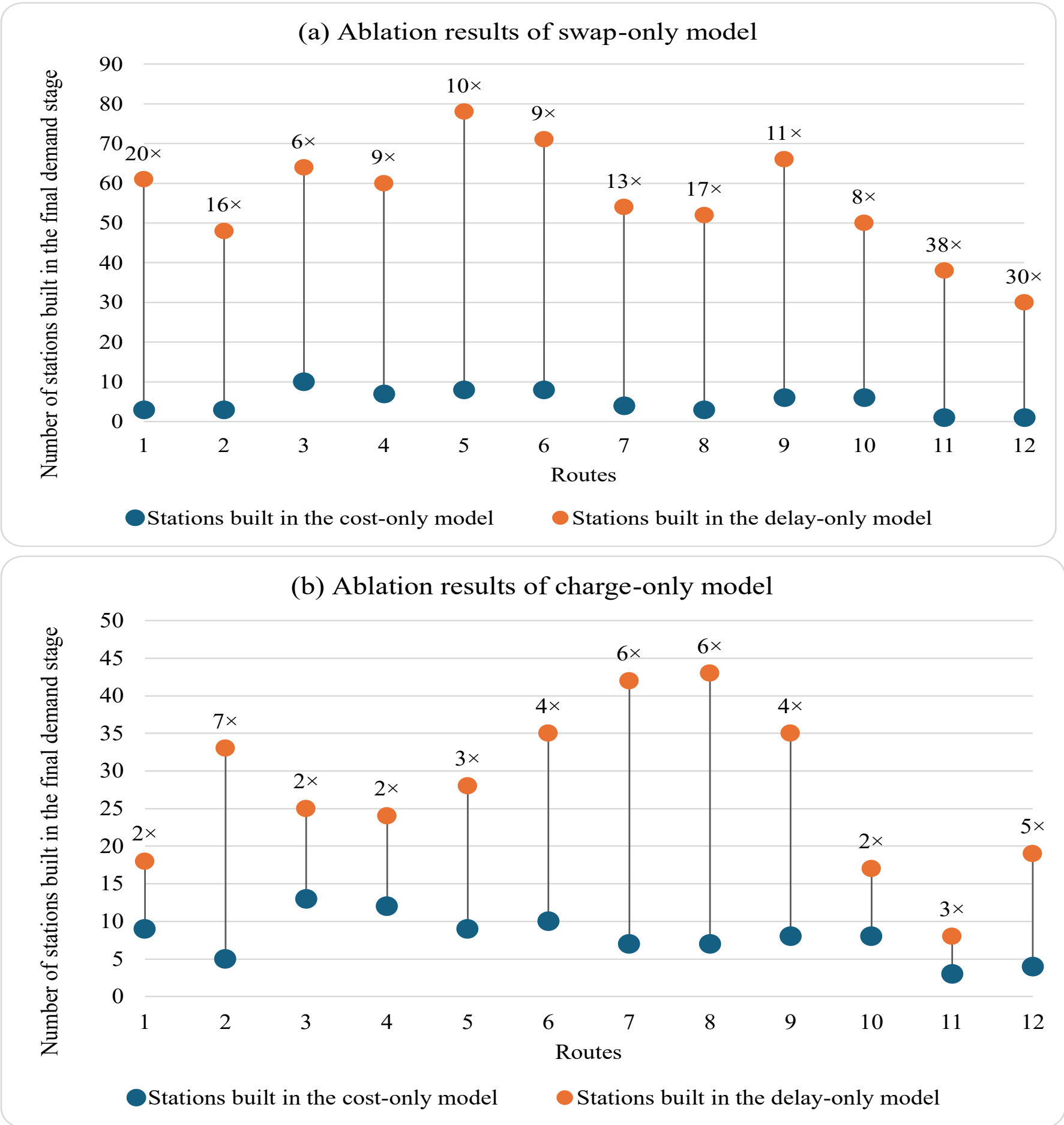

*Figure 6: Stations built in the final stage under cost-delay trade-off for swap-only (a) and charge-only (b)models.*

# 6 Discussion

The results in Section 5 bear on the three questions posed in the Introduction: (i) how much infrastructure the corridors require when energy, charging and rest are modelled explicitly; (ii) how sensitive that requirement is to the balance between capital cost and driver delay; and (iii) how much planning under adoption uncertainty changes which stations are built and when.

On the second question, the cost–delay trade-off is inactive at the assumed value of time. Comparing the combined-objective policies of Figure 4 with the cost-only policies (Figure 6) shows that the least-cost network satisfying the range, capacity and HoS constraints is already the cost–delay optimum on every charge-only route, whereas removing capital cost from the objective multiplies the station count by 2–7 for charging and opens nearly every candidate site for swapping. The delay-only extreme is therefore a diagnostic rather than a plausible investment target. The practical implication is that, once charging is aligned with mandatory rest, network size is governed by range and capacity feasibility at least capital cost rather than by delay. Whether this holds at higher values of time, or for drivers whose charging cannot count as rest, remains to be tested with a value-of-time sweep.

Comparing the deterministic rolling-horizon model versus the SDDiP model, the results illustrate the central methodological trade-off between the two approaches. Under the charge-only variant, the two approaches select similar total station counts and identical peak charger requirements in several stages, although the selected sites and early-stage costs differ. This comparison suggests that the value of the stochastic model is concentrated in investment timing and site choice under the assumed uncertainty, rather than in a uniform increase in station count. The swapping-only network differs more substantially from the deterministic benchmark. The SDDiP policy sometimes builds fewer but more concentrated battery swapping stations as demand grows, increasing the replacement inventory at the busiest locations. It also shifts part of the investment earlier because the policy accounts for possible higher-demand branches. In summary, the deterministic rolling-horizon model is fast, but has no way to hedge against the possibility that the realised demand differs from that path. The SDDiP model is far more computationally expensive to solve, but produces a policy that is explicitly optimised against the full range of demand uncertainty represented in the scenario tree, which is the more appropriate criterion for capital-intensive, long-lived infrastructure decisions of this kind.

The SDDiP model reached policy-evaluation confidence bounds of 5% or less on 19 of the 24 instances (combination of 12 routes and swapping/charging) despite the integer state space, though cost scales with route complexity, not iteration count. Per-iteration solve time varied by close to two orders of magnitude across routes, and the high-demand routes required both a significantly higher number of iterations and high-cost backward-pass dual solves per iteration. This suggests that computational budget for network-wide SDDiP deployment should be allocated by route complexity (candidate count, peak flow) rather than assumed uniform across a corridor set, an operational consideration for applying this framework at scale.

Overall, the experiments indicate that staged investment can produce different site and capacity trajectories from a deterministic rolling-horizon policy under the assumed demand process. They also show that terrain, charging-curve, rest and capacity assumptions can influence feasibility and delay. The magnitude of these effects should be established through systematic ablation and sensitivity experiments before the model is used for policy or investment decisions.

# 7 Conclusion

This study developed an integrated optimisation framework for siting battery-electric truck charging and swapping infrastructure that jointly addresses five gaps identified in the existing literature: terrain-derived, physics-based energy consumption in place of a flat distance-based rate; direct integration of mandatory driver HoS rest regulations into the charging/siting decision; multi-pack batteries modelled as an independently swappable set; a multi-stage stochastic SDDiP extension that times infrastructure investment under uncertain demand growth; and station siting that allows a route's peak-hour demand to be served jointly by multiple, capacity-contended facilities rather than a single site.

Applied to twelve major freight corridors on Australia's East Coast, using terrain data, real candidate infrastructure sites, and observed heavy-vehicle traffic volumes, the framework shows that (i) on the illustrative route every charging event fits within a mandatory break, and at the assumed value of time the least-cost network is also the cost–delay optimum; (ii) a ±30% change in adoption alters the final charge-only network by only −12% to +13% stations, with over 95% of stations built by the second stage; and (iii) planning under uncertainty mainly changes which sites are selected (71% overlap with a deterministic plan for charging and about half for swapping). The SDDiP policies reached statistical oconfidence bounds of 2.6–5.4% for swap-only and 3.1–7.9% for charge-only routes.

Methodologically, the SDDiP implementation combines several established elements and adds two results. It follows Zou et al. (2019) in generating Lagrangian cuts for the binary deployment state; it represents stagewise-dependent adoption by a Markov chain with cuts indexed by (stage, parent branch), as is standard for Markovian SDDP (Löhndorf and Shapiro, 2019); and it solves each Lagrangian dual with a level (bundle) method (Lemaréchal et al., 1995), which in preliminary tests converged where plain subgradient ascent did not. Proposition 1 shows that the cuts remain valid lower bounds when dual iterations or MILP solves are stopped early, so time limits weaken the cuts but never invalidate them, and Proposition 2 shows that the Lagrangian multipliers can be restricted to $[0, \rho C_i]$ without loss, which keeps the level method's master problem bounded. Finally, training is stopped on lower-bound stability alone and solution quality is reported from a separate, larger-sample policy evaluation, because with the small forward samples recommended by Zou et al. (2019) the per-iteration upper bound was too noisy to serve as a stopping signal.

This study, however, suffers from a few limitations, which set out the direction for future studies. First, the energy consumption is assumed as a function of terrain, vehicle characteristics, auxiliary loads and a fixed travel speed, and does not explicitly model traffic, weather, payload variation, or battery thermal effects on efficiency. In particular, the model relies on heavy vehicle traffic counts rather than payload data and therefore assumes a fixed vehicle mass across all trips; as Goeke and Schneider (2015) demonstrate, incorporating actual payload substantially improves energy-consumption accuracy, so more granular, trip-level payload data would be a natural refinement of the current approach.

Second, charging and swapping stations are assumed to have the same fixed establishment cost, deliberately excluding technology-specific equipment costs (chargers, battery inventory, automated swap systems) because these are highly uncertain, rapidly evolving, and vary considerably across manufacturers. Future studies can test technology-specific, time-varying cost curves as charging and swapping equipment pricing matures and becomes less volatile and publicly available to researchers.

Third, the transition probabilities and branch multipliers of the SDDiP model are hypothesised; they should be calibrated against historical BET uptake, fleet-operator surveys or forward order data once available. The proposed SDDiP model only carries the binary station-open decision forward, and charger counts and battery inventory are re-solved each stage without a ratchet constraint. Future studies could extend this by including the number of chargers and battery inventory with their own monotonicity constraint, so capacity expansions can be staged and irreversible. However, from the operational viewpoints and emerging business models, the economic viability of this option should be studied in more detail.

Fourth, the adoption-branch multiplier scales exogenous traffic volumes; the model does not capture the more realistic feedback where better charging/swapping infrastructure itself accelerates BET adoption (i.e. induced demand). Future studies can consider an endogenous demand model where next-stage adoption probabilities respond to the infrastructure actually built, not just to a pre-set Markov process.

Fifth, the hours-of-service rule is stylised, as all the routes in the case study had travel times less than 12 hours. Therefore, this study does not represent the 24-hour work limit and 7-hour stationary rest of the HVNL Standard Hours, which bind on routes longer than about 12 hours and would create overnight charging opportunities. Implementing the full sliding-window rules is a natural extension.

Sixth, each route is optimised separately, and the peak-hour flow is assumed to load every station on the route at the same time. A joint network model with time-shifted station arrivals would capture stations shared at corridor intersections and would likely reduce the capacity required.

Finally, this study overlooks the grid capacity and considers a fixed interconnection for the candidate sites without nearby EV charging stations. Future studies can integrate the station siting decision with the grid capacity constraints, heterogeneously introduced across candidate sites.

## Supplementary Material:

The model results are available through an interactive web-based dashboard at:
https://www.freightlabx.com/app/battery-truck-charging.